\documentclass[aps,pra,twocolumn,amssymb,amsfonts,floatfix,footinbib,superscriptaddress,longbibliography,notitlepage]{revtex4-1}
\usepackage[utf8]{inputenc}
\pdfoutput=1
\usepackage{color}
\usepackage{multirow}
\usepackage{dcolumn}
\usepackage{float}
\usepackage{xcolor}
\usepackage{amsmath,amsthm,amsfonts,amssymb}
\usepackage{times}
\usepackage[normalem]{ulem}
\usepackage{commath}
\usepackage{physics}
\usepackage{bm}
\usepackage[pdftex]{graphicx}
\usepackage[colorlinks=false]{hyperref}
\usepackage{enumitem}

\begin{document}

\title{Quantum Localization Measures in Phase Space}

\author{D. Villase\~nor}
\author{S. Pilatowsky-Cameo}
\affiliation{Instituto de Ciencias Nucleares, Universidad Nacional Aut\'onoma de M\'exico, Apdo. Postal 70-543, C.P. 04510 CDMX, Mexico}
\author{M. A. Bastarrachea-Magnani}
\affiliation{Departamento de F\'isica, Universidad Aut\'onoma Metropolitana-Iztapalapa, San Rafael Atlixco 186, C.P. 09340 CDMX, Mexico}
\author{S. Lerma-Hern\'andez}
\affiliation{Facultad de F\'isica, Universidad Veracruzana, Circuito Aguirre Beltr\'an s/n,  C.P. 91000 Xalapa, Veracruz, Mexico}
\author{J. G. Hirsch}
\affiliation{Instituto de Ciencias Nucleares, Universidad Nacional Aut\'onoma de M\'exico, Apdo. Postal 70-543, C.P. 04510  CDMX, Mexico}

%%%%%%%%%%%%%% ABSTRACT %%%%%%%%%%%%%%%%

\begin{abstract}

Measuring the degree of localization of quantum states in phase space is essential for the description of the dynamics and equilibration of quantum systems, but this topic is far from being understood. There is no unique way to measure localization, and individual measures can reflect different aspects of the same quantum state. Here, we present a general scheme to define localization in measure spaces, which is based on what we call {\it R\'enyi occupations}, from which any measure of localization can be derived. We apply this scheme to the four-dimensional unbounded phase space of the interacting spin-boson Dicke model. In particular, we make a detailed comparison of two localization measures based on the Husimi function in the regime where the model is chaotic, namely one that projects the Husimi function over the finite phase space of the spin and another that uses the Husimi function defined over classical energy shells. We elucidate the origin of their differences, showing that in unbounded spaces the definition of {\it maximal delocalization} requires a bounded reference subspace, with different selections leading to contextual answers.

\end{abstract}

%%%%%%%%%%%%%%%%%%%%%%%%%%%%%%%%%%%%%%%%
\maketitle
%%%%%%%%%%%%%%%%%%%%%%%%%%%%%%%%%%%%%%%%

%%%%%%%%%%%%%%%% INTRODUCTION %%%%%%%%%%%%%%%%

\section{INTRODUCTION}
\label{sec:Introduction}

The term  dynamical localization was coined to denote the  quantum limitation of classical diffusion in the chaotic regime~\cite{Chirikov1981,Izrailev1990}. Originally, it was first observed in periodically kicked rotors, and later it was also found in different systems, as the hydrogen atom in a monochromatic field and Rydberg atoms~\cite{Casati1984a,Casati1987,Blumel1987}, as well as related to the Anderson localization present in one-dimensional disordered systems~\cite{Anderson1958,Fishman1982}. The phenomenon was also observed  in conservative systems, such as the Band-random-matrix model and quantum billiards~\cite{Casati1993,Borgonovi1996,Batistic2010,Batistic2013a,Batistic2013b,Batistic2019,Robnik2020} and more recent studies have focused its onset on many-body systems~\cite{Rozenbaum2017a,Fava2020,Rylands2020}.

A usual way to measure delocalization is the exponential of an entropy~\cite{Campbell1966}, which, under different names, is widely used throughout different areas of science. For example, in ecology, the {\it diversity indices} are used to count the number of species in a population~\cite{Jost2006}. In information science and linguistics, the {\it perplexity} quantifies how well a probabilistic model fits some data~\cite{Jelinek1977,Brown1992}. In physics, localization --as the exponential of entropy-- is defined with respect to a given space. Using the phase space one may draw a connection between the structures of the classical dynamics with those of the quantum realm~\cite{Gorin1997}. Quantum states may be represented in the phase space through the so-called Husimi function~\cite{Husimi1940}, and the exponential of the Wehrl entropy~\cite{Wehrl1978}, which is the Shannon entropy of the Husimi function (or more generally R\'enyi-Wehrl entropies~\cite{Gnutzmann2001}), may be used to measure the localization of quantum states in that space.

Additionally to a particular entropy and space, specific subspaces or projections of the Husimi function may be necessary to choose in order to talk about maximally delocalized states. When the measure space is unbounded (i.e. has infinite volume), one may find states that are arbitrarily delocalized and needs to choose a region of finite volume which serves as a benchmark. We will see that there is no universal way to do this, and a series of fundamental choices which have direct repercussions on the behavior of the localization measures must me made.

Recent studies~\cite{Wang2020,Pilatowsky2021NatCommun} have used the exponential of the R\'enyi-Wehrl entropies to measure the localization  of eigenstates in the phase space of the Dicke model. This model is a collective many-body system~\cite{Dicke1954} initially introduced to explain the phenomenon of superradiance~\cite{Hepp1973a,Wang1973,Garraway2011,Kirton2019}. It describes a set of two-level atoms interacting collectively with a quantized radiation field and has been used extensively to study different phenomena, as out-of-time-ordered correlators~\cite{Chavez2019,Lewis-Swan2019,Pilatowsky2020}, quantum scarring~\cite{Deaguiar1992,Furuya1992,Bakemeier2013,Pilatowsky2021NatCommun,Pilatowsky2021}, and non-equilibrium dynamics~\cite{Altland2012NJP,Kloc2018,Lerma2018,Lerma2019,Kirton2019,Villasenor2020}. Some experimental realizations of this model involve superconducting circuits~\cite{Jaako2016}, cavity assisted Raman transitions~\cite{Baden2014,Zhang2018}, and trapped ions~\cite{Cohn2018,Safavi2018}.

A first purpose of this contribution is to identify  the general mathematical framework from which any measure of localization can be derived. We will call these measures, which are the exponential of the R\'enyi entropy, {\it R\'enyi volumes}~\cite{Hall1999}, and from them we will define the {\it R\'enyi occupations}, which are relative localization measures with respect to a finite reference volume. These general localization measures simplify to the usual generalized participation ratios for the case of a discrete set~\cite{OttBook,Murphy2011}.

In this work two R\'enyi occupations based on Husimi functions are explored in detail in the unbounded phase space of the Dicke model. The first one is defined in the atomic phase space of the model, where the reference volume is defined in the Bloch sphere~\cite{Wang2020}. The second R\'enyi occupation is defined over classical energy shells, where the reference volume corresponds to the volume of the classical energy shell~\cite{Pilatowsky2021NatCommun}. These two measures serve as examples for our general framework. We compare their behavior for eigenstates in the chaotic regime of the model and also their dynamical evolution for non-stationary states. Their dissimilar behaviors prove our assertion that there is no universal way to identify maximal delocalization in unbounded spaces.

The article is organized as follows. In Sec.~\ref{sec:LocalizationMeasures} we expose a general method to define localization measures in bounded or unbounded spaces, discrete or continuous. In Sec.~\ref{sec:DickeModel} we introduce the Dicke model and its classical limit. In Sec.~\ref{sec: OccMeasuresPhaseSpace} we focus on localization measures in the four-dimensional phase space of the Dicke model, constructing two different R\'enyi occupations. In Sec.~\ref{sec:PropertiesPhaseSpace} we compare the behaviors of both R\'enyi occupations for different kind of states, as eigenstates, evolved coherent states, as well as mixed coherent states in time and in space. Finally, our conclusions are presented in Sec.~\ref{sec:Conclusions}.

%%%%%%%%%%%%%%% SECTION II %%%%%%%%%%%%%%%%%%%%%%%%%

\section{R\'ENYI OCCUPATION IN GENERAL SPACES}
\label{sec:LocalizationMeasures}

Consider a space $X$ with a measure $\mathcal{V}$ that generates an integral $\int_X \dif \mathcal{V}(\bm x)\,\bullet$. This space may be discrete, in which case $\mathcal{V}(\Omega) =\abs{\Omega}$ ($\Omega\subseteq X$) is the counting measure, whose integral is just a sum $\int_X \dif \mathcal{V}(\bm x) \,\bullet =\sum_{\bm x\in X} \,\bullet$. Regardless of the space, we will call $\mathcal{V}(\Omega)=\int_\Omega \dif \mathcal{V}(\bm x)\in (0,\infty)$ the volume of $\Omega\subseteq X$. In $X$, we have normalized functions $\varphi\colon X\to [0,\infty)$ (i.e. $\int_X \dif \mathcal{V}(\bm x)\,\varphi(\bm x)=1$), which we will call  probability distributions~\footnote{If the space is discrete, then this probability density function  is just the probability at each point $\bm x_i \in X$, $p_i=\varphi(\bm x_i)$.}. 

For any  bounded measurable subset $\Omega\subseteq X$,  the uniform probability distribution on $\Omega$ is given by
\begin{equation}
\label{eq:unif_prob}
    \varphi_\Omega(\bm x)=\begin{cases}1/\mathcal{V}(\Omega)&\text{if }\bm x \in \Omega,\\ 0&\text{else.}\end{cases}
\end{equation}
We   define  the volume occupied by any probability distribution $\varphi$ in $X$, $\mathcal{V}(X,\varphi)$, by imposing the following properties:
\begin{itemize}
    \item The volume occupied by a uniform probability is $\mathcal{V}(X,\varphi_\Omega)=\mathcal{V}(\Omega)$.
    \item The volume is an homogeneous function of degree one under scaling of the measure $\mathcal{V}(\Omega)$, that is
    \begin{equation}
    \mathcal{V}(X',\varphi')=k\,\mathcal{V}(X,\varphi), \label{eq:scaling_property}
    \end{equation}
where $X'=X$ is the same space, but with volume element $\dif \mathcal{V}'=k\dif \mathcal{V}$, and  $\varphi'(\bm x)=\varphi(\bm x)/k$, with ${k>0}$. 
\end{itemize}
  
In App.~\ref{app:LocalizationMeasuresDetails}, we show that these two conditions naturally lead to the expressions
\begin{equation}
\label{eqn:OcuppiedVolumeOrderalpha}
\mathcal{V}_\alpha(X,\varphi)=\left(\int_X \dif \mathcal{V}(\bm x) \varphi(\bm x)^{\alpha}\right)^{1/(1-\alpha)},
\end{equation}
which are  the exponential of the R\'enyi entropy of order $\alpha\geq 0$, $H_\alpha(X,\varphi)=\log{}(\int_X \dif \mathcal{V}(\bm x)\, \varphi(\bm x)^{\alpha})/(1-\alpha)$~\cite{Renyi1961,Campbell1966}.
In the limit $\alpha\rightarrow 1$ we get
\begin{equation}
\label{eqn:OcuppiedVolumeOrder1}
\mathcal{V}_1(X,\varphi)=\exp(-\int_X \dif \mathcal{V}(\bm x) \varphi(\bm x)\log\varphi(\bm x)),
\end{equation}
which is the exponential of the Shannon entropy $H_1(X,\varphi)$~\cite{Shannon1948}. Due to this  close  relation with the R\'enyi entropies, we call $\mathcal{V}_\alpha(X,\varphi)$ the {\it R\'enyi volume} of order $\alpha$~\cite{Hall1999,Nath2020Arxiv}.

The numbers $\mathcal{V}_\alpha(X,\varphi)$ are always positive, and grow as $\varphi$ spreads more over $X$. For non-uniform distributions $\varphi$ the R\'enyi volume is the volume of the {\it effective} region occupied by $\varphi$, and its value depends strongly  on  $\alpha$. In any case, the R\'enyi volume  of $\varphi$  measures how delocalized is $\varphi$.

When the space $X$ is bounded, that is $\mathcal{V}(X)<\infty$, then $\mathcal{V}_\alpha(X,\varphi)\leq \mathcal{V}(X)$, and the maximum R\'enyi volume $\mathcal{V}_\alpha(X,\varphi)=\mathcal{V}(X)$ is attained by the uniform distribution over $X$. Moreover, if $\alpha \neq 0$, then $\mathcal{V}_\alpha(X,\varphi)=\mathcal{V}(X)$ occurs only for the uniform distribution.  Thus, for a bounded space $X$,  we can define the {\it R\'enyi occupation} of order $\alpha$ of $\varphi$ in $X$ through the ratio
\begin{equation}
\label{eqn:locmeasure}
 \mathfrak{L}_\alpha(X,\varphi)=\frac{\mathcal{V}_\alpha(X,\varphi)}{ \mathcal{V}(X)} \in (0,1].
\end{equation}

On the contrary, if $X$ is unbounded, that is $\mathcal{V}(X)=\infty$, we may find distributions $\varphi$ that are arbitrarily delocalized. This can be seen by considering $\Omega\subseteq X$ such that $\mathcal{V}(\Omega)$ is arbitrarily large and taking the uniform distribution in $\Omega$. Thus, an unbounded space does not allow to define a R\'enyi occupation directly. However, one may consider a smaller space $\widetilde X$ which is bounded, and transform the probability distributions from $X$ to $\widetilde X$. A general method to perform this involves three fundamental steps:
\begin{enumerate}[label={[\textbf{S\arabic*}]},leftmargin=*]
\item Choose a smaller space $\widetilde X$, which may be some region of $X$. \label{step:1}
\item  Choose a new volume element $\dd{ \widetilde{\mathcal{V}}}$ for $\widetilde X$, such that the total volume is finite $\widetilde{\mathcal{V}}(\widetilde{X})<\infty$.\label{step:2}
\item  Choose a transformation for the probability distributions  $\varphi\colon X \to [0,\infty)$ into probability distributions $\widetilde \varphi\colon \widetilde X \to [0,\infty)$. \label{step:3}
\end{enumerate}
By following these three steps one may use the Eq.~\eqref{eqn:locmeasure} with $\widetilde X$ and $\widetilde \varphi$ to define a R\'enyi occupation $\mathfrak{L}_\alpha(\widetilde X,\widetilde \varphi)$.

The fact that the exponential of an entropy provides a measure of localization~\cite{Campbell1966} should not be surprising: entropies are, by virtue of Boltzmann's entropy formula, the logarithm of the number of microstates that give rise to a certain macroscopic state. If we imagine $X$ to be the space of possible microstates of a system and $\varphi$ to be a distribution in $X$, then we can measure the level of delocalization of $\varphi$ in $X$ by just {\it counting} the number of microstates that compose it.

We close this section with a simple well-known example of the R\'enyi volume  $\mathcal{V}_\alpha$ found in discrete spaces. If $\mathcal{B}=\{\ket{\phi_k}\mid k\in N\subseteq \mathbb{N}\}$ is a basis of some Hilbert space of dimension $\abs{N}$, each state $\ket{\psi}$ defines the probability function $\varphi_\psi(\ket{\phi_k})=\abs{\braket{\psi}{\phi_k}}^2$. Setting $\mathcal{V}$ to be the counting measure, Eq.~\eqref{eqn:OcuppiedVolumeOrderalpha} becomes
\begin{equation}
\mathcal{V}_\alpha(\mathcal{B},\varphi_\psi)=\left(\sum_{k\in N} \abs*{\braket{\psi}{\phi_k}}^{2\alpha}\right)^{1/(1-\alpha)},    
\end{equation}
which correspond to the generalized quantum participation ratios~\cite{OttBook,Murphy2011}, where the case $\alpha~=~2$ reduces to
\begin{equation}
\mathcal{V}_{2}(\mathcal{B},\varphi_\psi)=P_R=\left(\sum_{k\in N} \abs{\braket{\psi}{\phi_k}}^{4}\right)^{-1},
\end{equation}
the standard quantum participation ratio, which has been used as a standard measure of the localization of state $\ket{\psi}$ in the basis $\mathcal{B}$~\cite{Bastarrachea2016PRE}.

%%%%%%%%%%%%%%% SECTION III %%%%%%%%%%%%%%%%%%%%%%%%%

\section{DICKE MODEL}
\label{sec:DickeModel}

In the next sections, we will study the R\'enyi occupations in the Dicke model, which describes the interaction between a set of two-level systems and a single-mode confined electromagnetic field~\cite{Dicke1954}. Setting $\hbar=1$, the Hamiltonian of the model can be written as
\begin{equation}
\label{eqn:qua_hamiltonian}
\hat{H}_{D}=\omega\hat{a}^{\dagger}\hat{a}+\omega_{0}\hat{J}_{z}+\frac{\gamma}{\sqrt{\mathcal{N}}}(\hat{J}_{+}+\hat{J}_{-})(\hat{a}^{\dagger}+\hat{a}),
\end{equation}
where $\omega$ is the radiation frequency of the electromagnetic field,  $\mathcal{N}$ is the number  of two-level atoms with  transition frequency $\omega_{0}$, and $\gamma$ is the atom-field coupling strength. $\hat{a}^{\dagger}$ ($\hat{a}$) is the bosonic creation (annihilation) operator of the field mode, $\hat{J}_{x,y,z}=(1/2)\sum_{k=1}^{\mathcal{N}}\hat{\sigma}_{x,y,z}^{k}$ are the collective pseudo-spin operators, and  $\hat{\sigma}_{x,y,z}$ are the Pauli matrices which satisfy the SU(2) algebra. $\hat{J}_{+}$ ($\hat{J}_{-}$) are the raising (lowering) collective pseudo-spin operator, defined by $\hat{J}_{\pm}=\hat{J}_{x}\pm i\hat{J}_{y}$.
 
The squared total pseudo-spin operator $\hat{\textbf{J}}^{2}=\hat{J}_{x}^{2}+\hat{J}_{y}^{2}+\hat{J}_{z}^{2}$ has eigenvalues $j(j+1)$, which specify different invariant subspaces of the model. In this work, we use the maximum pseudo-spin value $j=\mathcal{N}/2$, which defines the totally symmetric atomic subspace that includes the ground state.

The Dicke model develops a quantum phase transition when its coupling strength reaches the critical value $\gamma_{c}=\sqrt{\omega\omega_{0}}/2$~\cite{Hepp1973a,Hepp1973b,Wang1973,Emary2003}. At that point the system goes from a normal phase ($\gamma<\gamma_c$) to a superradiant phase ($\gamma>\gamma_c$). 

The model displays regular and chaotic behavior, depending on the Hamiltonian parameters and excitation energies~\cite{Chavez2016}. Here, we consider a coupling in the superradiant phase, $\gamma=2\gamma_c=1$, where the system is in the strong-coupling hard-chaos regime. Also, we choose the resonant frequency case $\omega=\omega_0=1$, and use rescaled energies to the system size $j=30$. The diagonalization techniques for the Dicke Hamiltonian are fully explained in App.~\ref{app:DiagonalizationBasis}.

\subsection{Classical Limit of the Dicke Model}
\label{subsec:DickeModel}

A classical Dicke Hamiltonian is obtained by taking the expectation value of the quantum Hamiltonian $\hat{H}_{D}$ under the tensor product of bosonic Glauber and atomic Bloch coherent states $|\bm x\rangle=|q,p\rangle\otimes~|Q,P\rangle$~\cite{Deaguiar1991,Deaguiar1992,Bastarrachea2014a,Bastarrachea2014b,Bastarrachea2015,Chavez2016,Villasenor2020}, and dividing it by the system size $j$, 
\begin{equation}
\begin{split}
h_\text{cl}(\bm x) & =\frac{\langle\bm{x}|\hat{H}_{D}|\bm{x}\rangle}{j} \\
 & =\frac{\omega}{2}\left(q^{2}+p^{2}\right)+\frac{\omega_{0}}{2}Z^{2}+2\gamma qQ\sqrt{1-\frac{Z^{2}}{4}}-\omega_{0},
\end{split}
\end{equation}
where $Z^{2}=Q^{2}+P^{2}$. The bosonic Glauber and atomic Bloch coherent states are, respectively,
\begin{equation}
\begin{split}
|q,p\rangle & =e^{-(j/4)\left(q^{2}+p^{2}\right)}e^{\left[\sqrt{j/2}\left(q+ip\right)\right]\hat{a}^{\dagger}}|0\rangle, \\
|Q,P\rangle & =\left(1-\frac{Z^{2}}{4}\right)^{j}e^{\left[\left(Q+iP\right)/\sqrt{4-Z^{2}}\right]\hat{J}_{+}}|j,-j\rangle,
\end{split}
\end{equation}
where $|0\rangle$ is the photon vacuum and $|j,-j\rangle$ is the state with all the atoms in the ground state.

The classical Hamiltonian $h_\text{cl}(\bm x)$ has a four-dimensional phase space $\mathcal{M}$ in the coordinates $\bm x=(q,p;Q,P)$. The rescaled classical energy $\epsilon=E/j$ that corresponds to $h_\text{cl}$ defines an effective Planck constant $\hbar_{\text{eff}}=1/j$~\cite{Ribeiro2006}.

%%%%%%%%%%%%%%% SECTION IV %%%%%%%%%%%%%%%%%%%%%%%%%

\section{R\'ENYI OCCUPATIONS IN THE PHASE SPACE OF THE DICKE MODEL}
\label{sec: OccMeasuresPhaseSpace}

\subsection{Husimi Function}
\label{subsubsec:HusimiFunction}

The Husimi function is a quasi-probability distribution function~\cite{Husimi1940,Hillery1984} defined as the expectation value of the density matrix $\hat{\rho}$ of an arbitrary state in the over-complete coherent-state basis $\{\ket{\bm x}\mid \bm x=(q,p;Q,P)\}$,
\begin{equation}
\label{eqn:Husimi}
\mathcal{Q}_{\hat{\rho}}(\bm x)= \langle \bm x | \hat{\rho} |\bm x \rangle.
\end{equation}
The Husimi function is a Wigner function~\cite{Wigner1932} smoothed by a Gaussian weight, and it is used to visualize how a state $\hat{\rho}$ is distributed in the phase space. In contrast to the Wigner function, the Husimi function is everywhere non-negative. When $\hat{\rho}=\dyad{\psi}$ is a pure state, it can be written as
\begin{equation}
\label{eqn:HusimiN}
\mathcal{Q}_{\psi}(\bm x)= | \langle \psi |\bm x \rangle |^{2}.
\end{equation}

\subsection{R\'enyi Volume in the Phase Space}
\label{subsubsec:PhaseSpace}

As a first step to study the R\'enyi occupations, we will calculate the R\'enyi volume for the phase space of the Dicke model $\mathcal{M}$ with the canonical volume element $\dif \mathcal{V}(\bm x)=\dif q\dif p\dif Q\dif P$, which we will denote by $\dif \bm x$. Each state $\hat{\rho}$ generates a probability distribution $\varphi_{\hat{\rho}}(\bm x)$ via the Husimi function [see Eq.~\eqref{eqn:Husimi}],
\begin{equation}
\varphi_{\hat{\rho}}(\bm x) =\frac{1}{C}\mathcal{Q}_{\hat{\rho}}(\bm x),
\end{equation}
where $C=\int_{\mathcal{M}}\dif \bm x \mathcal{Q}_{\hat{\rho}}(\bm x)=[2\pi/j]\cdot [4\pi/(2j+1)]$ ensures normalization.

By using Eqs.~\eqref{eqn:OcuppiedVolumeOrderalpha}~and~\eqref{eqn:OcuppiedVolumeOrder1} with $X={\mathcal{M}}$, $\varphi=\varphi_{\hat{\rho}}$, we get
\begin{equation}
\label{eqn:LocMeasurePhaseSpace1}
\mathcal{V}_1(\mathcal{M},\hat{\rho}) = C\exp(-\frac{1}{C}\int_\mathcal{M}  \dif \bm x \, \mathcal{Q}_{\hat{\rho}}(\bm{x})\log\mathcal{Q}_{\hat{\rho}}(\bm{x})),
\end{equation}
and
\begin{equation}
\label{eqn:LocMeasurePhaseSpacealpha}
\mathcal{V}_\alpha(\mathcal{M},\hat{\rho}) = C^{\alpha/(\alpha-1)}\left( \int_\mathcal{M}  \dif \bm x \, \mathcal{Q}_{\hat{\rho}}^{\alpha}(\bm{x})\right)^{1/(1-\alpha)}
\end{equation}
for $\alpha\geq 0$. Equation~\eqref{eqn:LocMeasurePhaseSpace1} is the exponential of the Wehrl entropy~\cite{Wehrl1978}. These types of measures have been studied in the SU(2) two-dimensional phase space in the Refs.~\cite{Gnutzmann2001,Goldberg2020}.

By definition, $\mathcal{V}_\alpha(\mathcal{M},\hat{\rho})>0$, but the quantum uncertainty principle actually provides a positive lower bound on the localization of a quantum state in the phase space. Namely, if $\alpha \gg \hbar_\text{eff}^2 $,
\begin{equation}
\label{eq:locMeasureSmallerThath}
    \mathcal{V}_\alpha(\mathcal{M},\hat{\rho})\geq(2 \pi  \hbar_\text{eff})^2 \alpha^\frac{2}{\alpha-1},
\end{equation}
where $\hbar_\text{eff}=1/j$ (see App.~\ref{app:LowerBoundOnPhaseSpaceLocalization} for details). Nevertheless, there is no upper bound on the R\'enyi volume occupied by a quantum state because  the phase space of the Dicke model is unbounded,
$\mathcal{V}(\mathcal{M})=\int_\mathcal{M}\dif{ \bm x}=\infty$. Thus, we can find arbitrarily delocalized states.

In order to define a R\'enyi occupation in the phase space of the Dicke model $\mathcal{M}$, we may apply the procedure~\ref{step:1}-\ref{step:3} outlined in Sec.~\ref{sec:LocalizationMeasures} for unbounded spaces. This is done by restricting the measure to bounded regions of $\mathcal{M}$. We will focus on two different R\'enyi occupations for $\mathcal{M}$, which use the atomic subspace and the classical energy shells as bounded regions.

\subsection{R\'enyi Occupation in the Atomic Subspace}
\label{subsubsec:AlternativeLocMeasures}

A R\'enyi occupation using the atomic subspace of the Dicke model was originally studied in the Ref.~\cite{Wang2020}. The complete phase space of the Dicke model is conformed of a bounded (atomic) subspace and an unbounded (bosonic) subspace. In some sense, selecting the atomic subspace as the finite-volume reference region seems to be the most natural choice to construct a R\'enyi occupation. This can be done by following the steps~\ref{step:1}-\ref{step:3} of Sec.~\ref{sec:LocalizationMeasures} (see Table~\ref{tab:01} for a summary of the steps). First, we choose the atomic subspace 
\begin{flalign}
    [\textbf{S1}] &&  \mathcal{A}=\{(Q,P) \mid Q^2 + P^2\leq 4\},&&
\end{flalign}
with the canonical area element
\begin{flalign}
   [\textbf{S2}]&& \dif \mathcal{V}(Q,P)=\dif Q\dif P.&&
\end{flalign}
The subspace $\mathcal{A}$ is bounded with respect to $\mathcal{V}$,
\begin{flalign}
      \mathcal{V}(\mathcal{A})=\int_{\mathcal{A}} \dif Q\dif P=4\pi.
\end{flalign}

For each state $\hat{\rho}$ consider the probability distribution given by the projection of the Husimi function into ${\mathcal{A}}$,
\begin{flalign}
[\textbf{S3}]&&\varphi_{{\mathcal{A}},\hat{\rho}}(Q,P)=\frac{1}{C}\widetilde{\mathcal{Q}}_{\hat{\rho}}(Q,P),&&
\end{flalign}
where
\begin{equation}
\label{eqn:Husimiprojection}
\widetilde{\mathcal{Q}}_{\hat{\rho}}(Q,P)=\iint\dif q\dif p\, \mathcal{Q}_{\hat{\rho}}(q,p;Q,P)
\end{equation}
and $C=[2\pi/j]\cdot [4\pi/(2j+1)]$ ensures normalization. Using Eq.~\eqref{eqn:locmeasure} with $X={\mathcal{A}}$, $\varphi=\varphi_{{\mathcal{A}},\hat{\rho}}$, and $\mathcal{V}(\mathcal{A})=4\pi$, we obtain the R\'enyi occupations on the atomic subspace
\begin{multline}
\label{eqn:RobnikMeasures1}
\mathfrak{L}_1({\mathcal{A}},\hat{\rho})=\\
\frac{C}{4\pi}\exp\left(-\frac{1}{C}\int_{\mathcal{A}}  \dif Q \dif P \,  \widetilde{\mathcal{Q}}_{\hat{\rho}}(Q,P) \log \widetilde{\mathcal{Q}}_{\hat{\rho}}(Q,P)\right),
\end{multline}
and
\begin{equation}
\label{eqn:RobnikMeasuresalpha}
\mathfrak{L}_{\alpha}({\mathcal{A}},\hat{\rho}) = \frac{C^{\alpha/(\alpha-1)}}{4\pi}\left( \int_{\mathcal{A}}  \dif Q \dif P \, \widetilde{\mathcal{Q}}_{\hat{\rho}}^{\alpha}(Q,P)\right)^{1/(1-\alpha)}.
\end{equation}

The R\'enyi occupation of order $\alpha=2$, which we will henceforth focus on, reads 
\begin{equation}
\label{eqn:RobnikMeasures2}
\mathfrak{L}_2({\mathcal{A}},\hat{\rho}) = \frac{C^2}{4\pi}\left( \int_{\mathcal{A}}  \dif Q \dif P \, \widetilde{\mathcal{Q}}_{\hat{\rho}}^{2}(Q,P)\right)^{-1},
\end{equation}
which along with that of order $\alpha=1$, was studied in detail in the Ref.~\cite{Wang2020} for eigenstates $\hat{\rho}=\dyad{E_k}$ of the Dicke model. In that work, these R\'enyi occupations were also multiplied by an additional factor which equals the percentage of $\mathcal{A}$ that is covered by classical chaotic trajectories, and a linear relation between $\mathfrak{L}_1({\mathcal{A}},\hat{\rho})$ and $\mathfrak{L}_2({\mathcal{A}},\hat{\rho})$ was found for some of the eigenstates in the chaotic regime.

%%%%%%%%%%%%%%%%%%%% FIGURE 1 %%%%%%%%%%%%%%%%%%%%%%%%%%%

\begin{figure*}[ht]
\centering
\includegraphics[width=0.95\textwidth]{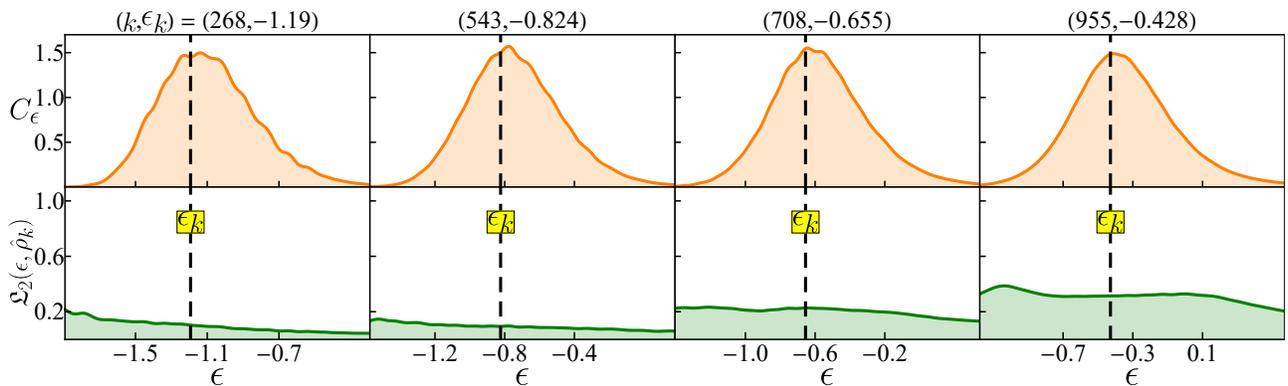}
\caption{\textbf{Top panels:} Energy profile, $C_{\epsilon}$, of the Husimi function over the classical energy shell at $\epsilon$ for selected eigenstates $\hat{\rho}_{k}$ with $k=268,543,708,955$ in the energy region $\epsilon_{k}\in(-1.2,-0.43)$. \textbf{Bottom panels:} R\'enyi occupation $\mathfrak{L}_{2}(\epsilon,\hat{\rho}_{k})$ [see Eq.~\eqref{eqn:LocMeasureEnergyShell2}] as a function of the energy $\epsilon$ for the same selected eigenstates. In both the top and bottom panels, the vertical dashed black lines indicate the eigenenergy $\epsilon_{k}$ of the corresponding eigenstate $\hat{\rho}_{k}$. The system size is $j=30$.  
}
\label{fig01}
\end{figure*}

%%%%%%%%%%%%%%%%%%%%%%%%%%%%%%%%%%%%%%%%%%%%%%%%%%%%%

\subsection{R\'enyi Occupation over Classical Energy Shells}
\label{subsubsec:LocMeasureEnergyShell}

Because of the energy conservation, the temporal evolution of any initial state under a time-independent Hamiltonian will maintain the same average energy. States with a well-defined energy center will remain close to the corresponding  classical energy shell in the phase space.
 
This motivates us to define another R\'enyi occupation by using the bounded classical energy shells as reference subspaces, as it was done in the Ref.~\cite{Pilatowsky2021NatCommun}. This can be achieved by applying the steps~\ref{step:1}-\ref{step:3} of the Sec.~\ref{sec:LocalizationMeasures} as follows (see Table~\ref{tab:01} for a summary). 

Consider the classical energy shell at energy $\epsilon$ 
\begin{flalign}
\label{eqn:SubspaceShell}
[\textbf{S1}] &&
\mathcal{M}_\epsilon=\{\bm x=(q,p;Q,P)\mid h_\text{cl}(\bm{x})=\epsilon \},&&
\end{flalign}
with the surface volume element
\begin{flalign}
\label{eq:MeasureInEnergyShell}
[\textbf{S2}]&&
    \dif\mathcal{V}(\bm x) =\delta( h_\text{cl}(\bm{x}) - \epsilon) \dif \bm x,&&
\end{flalign}
which we will denote by $\dif \bm s$. The subspace $\mathcal{M}_\epsilon$ is bounded with respect to $\mathcal{V}$. In fact, the finite value  $\mathcal{V}(\mathcal{M}_\epsilon)$ can be calculated through the quantity
\begin{equation}
\label{eq:VolAndDensityOfStates}
    \nu(\epsilon)=\frac{1}{4\pi^2}{\mathcal{V}(\mathcal{M}_\epsilon)}=\frac{1}{4\pi^2}\int_\mathcal{M_\epsilon}\!\!\!\!\dif \bm s,
\end{equation}
which is the lowest order semmiclassical approximation for the quantum density of states obtained through the Gutzwiller trace formula \cite{Gutzwiller1971,Gutzwiller1990book}. See Ref.~\cite{Bastarrachea2014a} for a detailed analytical expression of this semiclassical density of states $\nu(\epsilon)$. 

For each state $\hat{\rho}$, consider the probability distribution given by the restriction of the Husimi function to $\mathcal{M}_\epsilon$,
\begin{flalign}
[\textbf{S3}]&&
\label{eq:EnergyShellProbabilityDistribution}
\varphi_{\epsilon,\hat{\rho}}(\bm x)=\frac{1}{C_\epsilon}\mathcal{Q}_{\hat{\rho}}(\bm x),    &&
\end{flalign}
where $C_\epsilon=\int_{\mathcal{M}_\epsilon}\dif\bm s \mathcal{Q}_{\hat{\rho}}(\bm x)$ ensures normalization.
Using Eq.~\eqref{eqn:locmeasure} with $X=\mathcal{M}_\epsilon$, $\varphi=\varphi_{\epsilon,\hat{\rho}}$, and $\mathcal{V}(\mathcal{M}_\epsilon)=4\pi^2\nu(\epsilon)$, we obtain the set of R\'enyi occupations over the energy shell at $\epsilon$,
\begin{equation}
\label{eqn:LocMeasureEnergyShell1}
\mathfrak{L}_1(\epsilon,\hat{\rho}) = \frac{C_\epsilon}{4\pi^2\nu(\epsilon)}\exp(-\frac{1}{C_\epsilon}\int_{\mathcal{M}_\epsilon}\!\!\!\!  \dif \bm s \,  \mathcal{Q}_{\hat{\rho}}(\bm{x})\log\mathcal{Q}_{\hat{\rho}}(\bm{x})),
\end{equation}
and
\begin{equation}
\label{eqn:LocMeasureEnergyShellalpha}
\mathfrak{L}_\alpha(\epsilon,\hat{\rho}) = \frac{C_\epsilon^{\alpha/(\alpha-1)}}{4\pi^2\nu(\epsilon)}\left( \int_{\mathcal{M}_\epsilon}  \!\!\!\!\dif \bm s \, \mathcal{Q}_{\hat{\rho}}^{\alpha}(\bm{x})\right)^{1/(1-\alpha)}.
\end{equation}

The R\'enyi occupation of order $\alpha=2$ 
\begin{equation}
\label{eqn:LocMeasureEnergyShell2}
\mathfrak{L}_2(\epsilon,\hat{\rho}) = \frac{C_\epsilon^2}{4\pi^2\nu(\epsilon)}\left( \int_{\mathcal{M}_\epsilon}\!\!\!\!  \dif \bm s \, \mathcal{Q}_{\hat{\rho}}^{2}(\bm{x})\right)^{-1},
\end{equation}
was used in the Ref.~\cite{Pilatowsky2021NatCommun}, to measure localization of eigenstates and temporal averaged evolved states in the Dicke model in relation to the phenomenon of quantum scarring. It was also used as a measure of quantum ergodicity. Henceforth, we will focus on this specific case $\alpha=2$ and leave a detailed study of the dependence on the order $\alpha$ and its relation with multifractality~\cite{OttBook,Schreiber1991,Mirlin2000,Mirlin2000b,Evers2008,Rodriguez2010,Martin2010,Dubertrand2015} for a future work.

Given a quantum state $\hat{\rho}$ with mean energy $\epsilon_{\hat{\rho}}=\tr\big(\hat{\rho} \hat{H}_D\big)$, we may measure its localization in the phase-space energy shell at any energy  $\epsilon$ through $\mathfrak{L}_2(\epsilon,\hat{\rho})$. Nevertheless, for eigenstates $\hat{\rho}_k=\dyad{E_k}$, a  natural  choice is  $\epsilon=\epsilon_{\hat{\rho}_k}=\epsilon_k$. Moreover, $\mathfrak{L}_2(\epsilon,\hat{\rho}_k)$ remains almost constant for the  energies around $\epsilon_k=E_k/j$ that are significantly populated by the state $\hat{\rho}_{k}$. The upper panels of Fig.~\ref{fig01}. show the energy profiles, $C_\epsilon=\int_{\mathcal{M}_\epsilon}\dif\bm s \mathcal{Q}_{\hat{\rho}}(\bm x)$, of four selected eigenstates, one ($k=268$) located in a mixed energy region where coexist regularity and chaos, and three ($k=543,708,955$) located in the fully chaotic energy regime. These energy profiles show that the Husimi functions of the respective eigenstates are concentrated in energy shells with relevant values in an interval around $\epsilon\sim \epsilon_k$, while the bottom panels show that $\mathfrak{L}_2(\epsilon,\hat{\rho}_k)$ is almost constant in the region of significant population. It is worth mentioning that we have found that the R\'enyi occupation $\mathfrak{L}_2(\epsilon,\hat{\rho}_k)$ can be more sensitive to $\epsilon$  at low energies due to the regularity of the dynamics and because at energies close to the ground-state energy $\epsilon_\text{GS}=-2.125$, the classical energy shells are very small, and their size increases rapidly as energy does. Nevertheless, for the energy regime studied in this work, these effects are negligible.

\begin{table}
\begin{tabular}{|c|c|c|c|c|}
\hline\hline 
\quad & \textbf{S1} & \multicolumn{2}{c|}{\textbf{S2}} & \textbf{S3} \\
\hline
  R\'enyi & Subspace  & Volume element & Volume  & Distribution \\
    occupation &  $\widetilde{X}$ & $\dif\mathcal{V}$ &  $\mathcal{V}(\widetilde{X})$ & 
    $\varphi_{\widetilde{X},\hat{\rho}}$ \\[0.5ex]
    \hline\hline
    & & & &\\[-1ex]
 $\displaystyle\mathfrak{L}_\alpha({\mathcal{A}},\hat{\rho})$ & $\displaystyle\mathcal{A}$ & $\dif Q\dif P$ & $\displaystyle4\pi$ &  $\displaystyle\frac{\widetilde{\mathcal{Q}}_{\hat{\rho}}(Q,P)}{C}$  \\[2ex]   \hline & & & &\\[-1ex]
 $\displaystyle\mathfrak{L}_\alpha(\epsilon,\hat{\rho})$ & $\displaystyle\mathcal{M}_\epsilon$ & $\delta(h_\text{cl}(\bm x)-\epsilon)\dif \bm x$ & $\displaystyle4\pi^2\nu(\epsilon)$ &  $\displaystyle\frac{\mathcal{Q}_{\hat{\rho}}(\bm x)}{C_\epsilon}$
 \\[2ex]  
 \hline\hline
\end{tabular}
\caption{\label{tab:01}
Fundamental steps made in the construction of the R\'enyi occupations of order $\alpha$ (see Sec.~\ref{sec:LocalizationMeasures}), defined in the atomic subspace of the Dicke model $\mathfrak{L}_\alpha(\mathcal{A},\hat{\rho})$ and over classical energy shells $\mathfrak{L}_\alpha(\epsilon,\hat{\rho})$.}
\end{table}

%%%%%%%%%%%%%%%%%%%% FIGURE 2 %%%%%%%%%%%%%%%%%%%%%%%%%%%

\begin{figure*}[ht]
\centering
\includegraphics[width=0.95\textwidth]{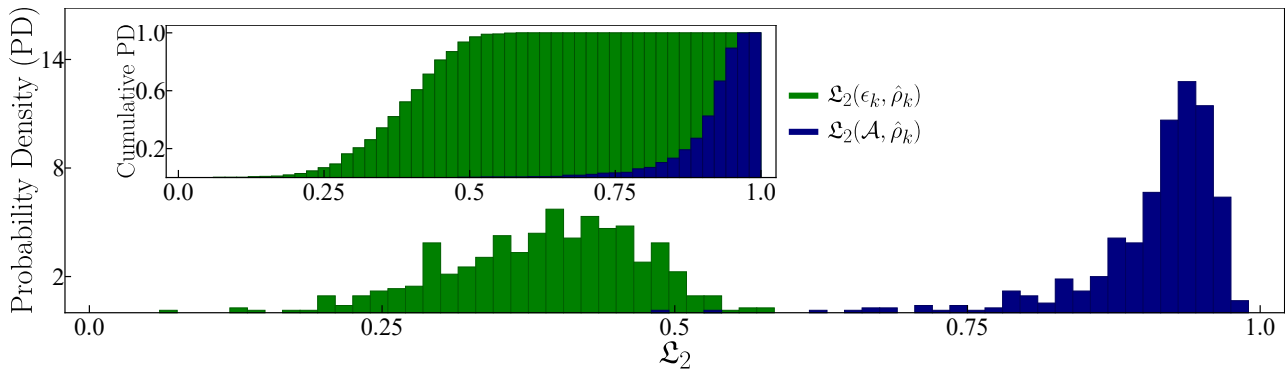}
\caption{Statistical distributions of the R\'enyi occupations  $\mathfrak{L}_{2}(\mathcal{A},\hat{\rho}_{k})$ (blue bars) and $\mathfrak{L}_{2}(\epsilon_{k},\hat{\rho}_{k})$ (green bars) [see Eqs.~\eqref{eqn:RobnikMeasures2}~and~\eqref{eqn:LocMeasureEnergyShell2}] for a set of 501 eigenstates $\hat{\rho}_{k}$ with $k\in[3121,3621]$, located in the chaotic-energy region $\epsilon_{k}\in(1.0,1.274)$. The inset shows their corresponding cumulative distributions. The system size is $j=30$.  
}
\label{fig02}
\end{figure*}

%%%%%%%%%%%%%%%%%%%%%%%%%%%%%%%%%%%%%%%%%%%%%%%%%%%%%

%%%%%%%%%%%%%%% SECTION V %%%%%%%%%%%%%%%%%%%%%%%%%

\section{PROPERTIES OF R\'eNYI OCCUPATIONS IN THE PHASE SPACE OF THE DICKE MODEL}
\label{sec:PropertiesPhaseSpace}

The R\'enyi occupations that we presented in the previous section have been used to measure the localization of the eigenstates in the chaotic regime~\cite{Wang2020,Pilatowsky2021NatCommun}. The statistical properties of the localization of eigenstates can be tied to the chaoticity of quantum systems. For example, a relatively low value of R\'enyi occupation can be a signal of quantum scarring. The localization of the eigenstates also has an impact in the evolution of non-stationary states, such as coherent states. Thus, studying the dynamical evolution of the localization of coherent states also provides very useful information about the dynamical properties of the system.

In this section we compare the R\'enyi occupations of the Husimi projection over the atomic subspace $\mathfrak{L}_2(\mathcal{A},\hat{\rho})$ [see Eq.~\eqref{eqn:RobnikMeasures2}] and over classical energy shells $\mathfrak{L}_2(\epsilon,\hat{\rho})$ [see Eq.~\eqref{eqn:LocMeasureEnergyShell2}], for eigenstates, time evolved coherent states, and coherent states mixed in phase space.

\subsection{Localization of Eigenstates}
\label{subsec:LocEigenstates}

For a set of 501 eigenstates of the Dicke model $\hat{\rho}_{k}=|E_{k}\rangle\langle E_{k}|$ with $k\in[3121,3621]$, located in the chaotic-energy region $\epsilon_k\in(1,1.274)$, we compute both R\'enyi occupations $\mathfrak{L}_{2}(\mathcal{A},\hat{\rho}_{k})$ and $\mathfrak{L}_{2}(\epsilon_k,\hat{\rho}_{k})$. The distributions for both occupations are shown in Fig.~\ref{fig02}. In this figure we see that the occupation $\mathfrak{L}_{2}(\mathcal{A},\hat{\rho}_k)$ clusters around the mean value $\mathfrak{L}_{2}\sim0.9$, which means that the Husimi projections  of all eigenstates in this chaotic region are almost completely delocalized in the atomic Bloch sphere $\mathcal{A}$. On the other hand, the occupation $\mathfrak{L}_{2}(\epsilon_k,\hat{\rho}_k)$ clusters around a mean value $\mathfrak{L}_{2}\sim0.4$, indicating that all eigenstates occupy less that half of the classical energy shell at $\epsilon_k$.

The striking differences between both R\'enyi occupations $\mathfrak{L}_{2}(\mathcal{A},\hat{\rho}_{k})$ and $\mathfrak{L}_{2}(\epsilon_k,\hat{\rho}_{k})$, indicate that they gauge different aspects of the same eigenstates, and show that one has to be cautious to interpret the values provided by different localization measures. Below we provide several numerical tests allowing to clarify the origin of the differences obtained for both R\'enyi occupations.

%%%%%%%%%%%%%%%%%%%% FIGURE 3 %%%%%%%%%%%%%%%%%%%%%%%%%%%

\begin{figure*}[ht]
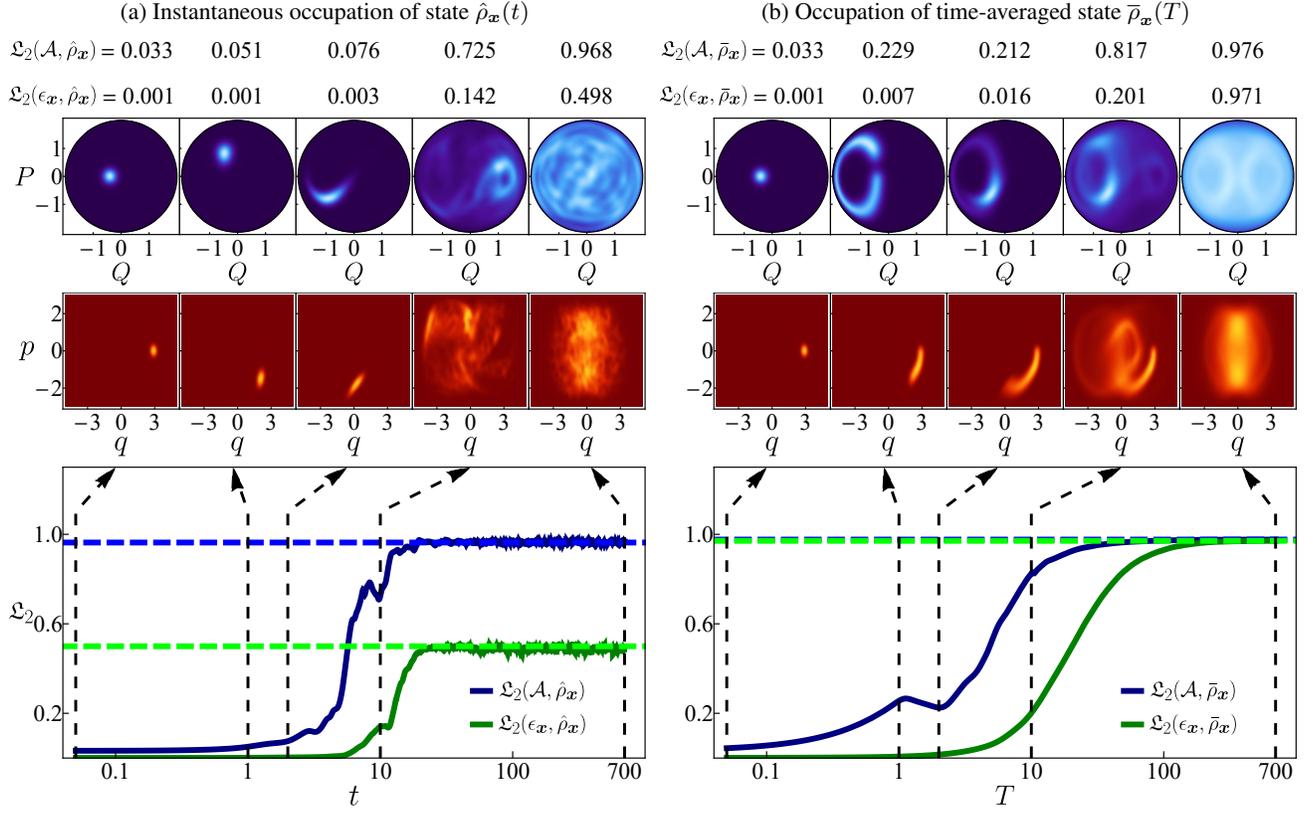

\centering
\begin{tabular}{cc}
(a) Instantaneous occupation of state $\hat{\rho}_{\bm{x}}(t)$  & (b) Occupation of time-averaged state $\overline{\rho}_{\bm{x}}(T)$ \\
\includegraphics[width=0.475\textwidth]{fig3a.pdf} &
\includegraphics[width=0.475\textwidth]{fig3b.pdf}
\end{tabular}
\caption{\textbf{Top squared panels:} Husimi function projected in the atomic coordinate plane $(Q,P)$ at different times $t$ and $T$ for both pure $\hat{\rho}_{\bm{x}}(t)$ (a) and time-mixed $\overline{\rho}_{\bm{x}}(T)$ (b) coherent states [see Eqs.~\eqref{eqn:pureCS}~and~\eqref{eqn:mixedCS}]. \textbf{Bottom rectangular panels:} R\'enyi occupations  $\mathfrak{L}_2(\mathcal{A},\hat{\rho}_{\bm{x}}(t))$ [$\mathfrak{L}_2(\mathcal{A},\overline{\rho}_{\bm{x}}(t))$] (solid blue curve) and $\mathfrak{L}_2(\epsilon_{\bm{x}},\hat{\rho}_{\bm{x}}(t))$ [$\mathfrak{L}_2(\epsilon_{\bm{x}},\overline{\rho}_{\bm{x}}(t))$] (solid green curve) [see Eqs.~\eqref{eqn:RobnikMeasures2}~and~\eqref{eqn:LocMeasureEnergyShell2}], for a pure initial coherent state $\hat{\rho}_{\bm{x}}(t)$ (a) and time-mixed coherent states $\overline{\rho}_{\bm{x}}(T)$ (b) in the chaotic-energy region $\epsilon_{\bm{x}}=1$. The selected initial coherent state $\hat{\rho}_{\bm{x}}$ is defined by the phase-space coordinates $\bm{x}=(2.894,0;-0.4,0)$ with energy width $\sigma_{\bm{x}}=0.693$ (units of $\epsilon$). Horizontal dashed light (green and blue) lines indicate the asymptotic value of each measure in both panels. Vertical dashed black lines indicate the value of $t$ and $T$ where the Husimi projections are shown. The system size is $j=30$.
}
\label{fig03}
\end{figure*}

%%%%%%%%%%%%%%%%%%%%%%%%%%%%%%%%%%%%%%%%%%%%%%%%%%%%%

%%%%%%%%%%%%%%%%%%%% FIGURE 4 %%%%%%%%%%%%%%%%%%%%%%%%%%%

\begin{figure*}[ht]
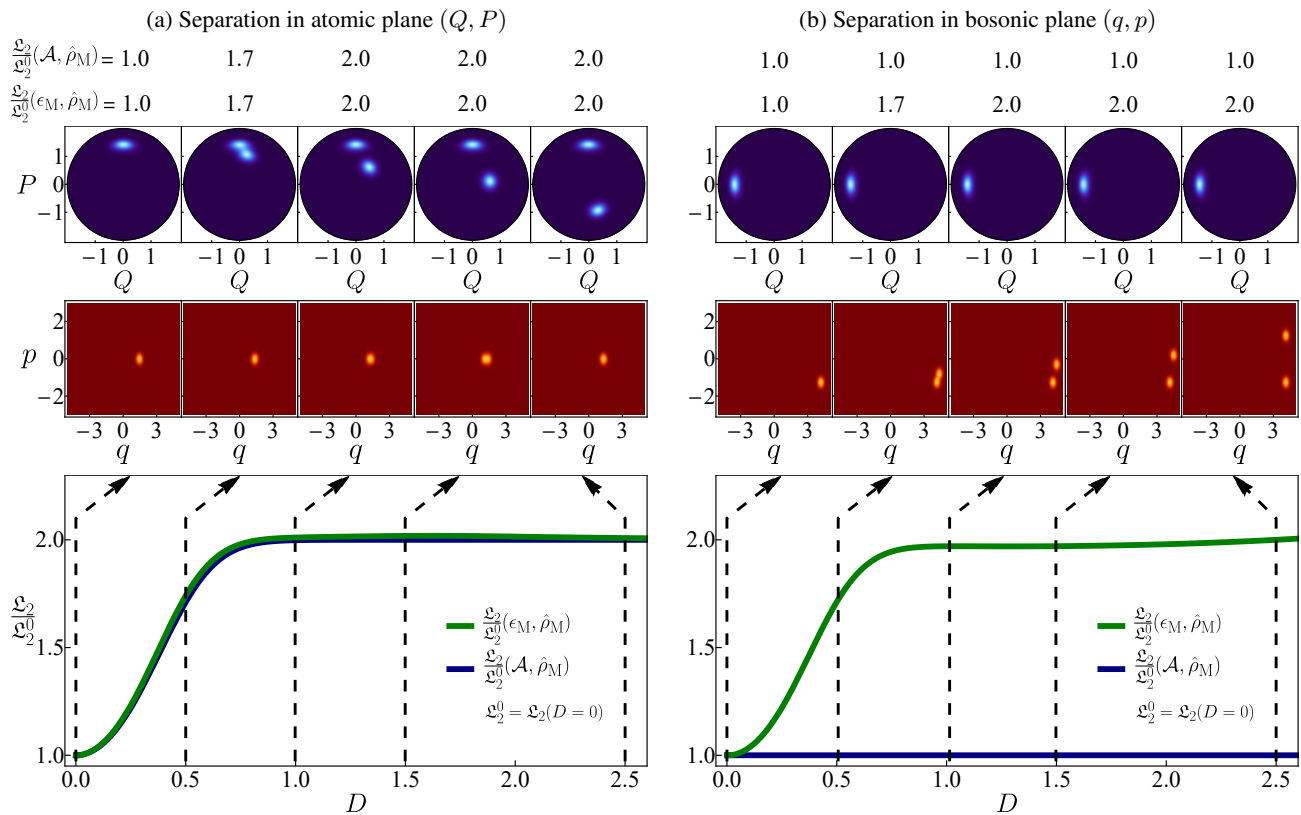

\centering
\begin{tabular}{cc}
(a) Separation in atomic plane $(Q,P)$ & (b) Separation in bosonic plane $(q,p)$ \\
\includegraphics[width=0.475\textwidth]{fig4a.pdf} &
\includegraphics[width=0.475\textwidth]{fig4b.pdf}
\end{tabular}
\caption{\textbf{Top squared panels:} Husimi function projected in the atomic $(Q,P)$ (blue circles) and bosonic $(q,p)$ (red squares) coordinate planes at different phase-space separations $D$. Panels in (a) depict the separation in the atomic plane $(Q,P)$, while panels in (b) in the bosonic one $(q,p)$. \textbf{Bottom rectangular panels:} R\'enyi occupations $\mathfrak{L}_2(\mathcal{A},\hat{\rho}_{\text{M}})/\mathfrak{L}_{2}^{0}$ (solid blue curve) and $\mathfrak{L}_2(\epsilon_{\text{M}},\hat{\rho}_{\text{M}})/\mathfrak{L}_{2}^{0}$ (solid green curve) rescaled to their initial values $\mathfrak{L}_{2}^{0}=\mathfrak{L}_{2}(D=0)$ [see Eqs.~\eqref{eqn:RobnikMeasures2}~and~\eqref{eqn:LocMeasureEnergyShell2}] as a function of the phase-space separation $D$, for mixed coherent states $\hat{\rho}_{\text{M}}(D)$ [see Eq.~\eqref{eqn:mixedCSD}] with constant energy width $\sigma=0.429$ (a) and $\sigma=0.342$ (b) [units of $\epsilon$]  in the chaotic-energy region $\epsilon_{\text{M}}=1$. Vertical dashed black lines indicate the value of $D$ where the Husimi projections are shown. The system size is $j=30$.
}
\label{fig04}
\end{figure*}

%%%%%%%%%%%%%%%%%%%%%%%%%%%%%%%%%%%%%%%%%%%%%%%%%%%%%

%%%%%%%%%%%%%%%%%%%% FIGURE 5 %%%%%%%%%%%%%%%%%%%%%%%%%%%

\begin{figure*}[ht]
\centering
\begin{tabular}{c}
Saturation of atomic plane $(Q,P)$\\
\includegraphics[width=0.95\textwidth]{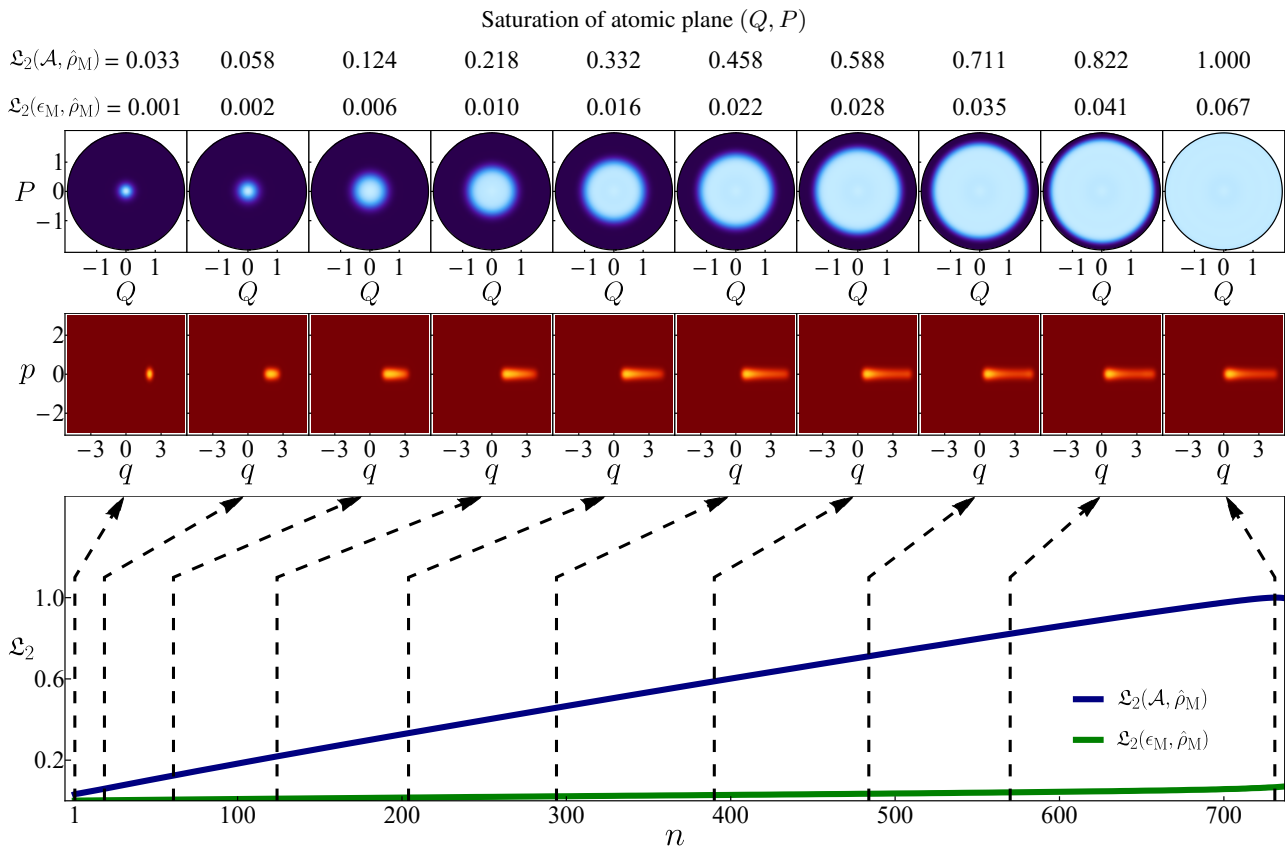}
\end{tabular}
\caption{\textbf{Top squared panels:} Husimi function projected in the atomic $(Q,P)$ (blue circles) and bosonic $(q,p)$ (red squares) coordinate planes at different number of added states $n$. \textbf{Bottom rectangular panel:} R\'enyi occupations $\mathfrak{L}_{2}(\mathcal{A},\hat{\rho}_{\text{M}})$ (solid blue curve) and $\mathfrak{L}_{2}(\epsilon_{\text{M}},\hat{\rho}_{\text{M}})$ (solid green curve) [see Eqs.~\eqref{eqn:RobnikMeasures2}~and~\eqref{eqn:LocMeasureEnergyShell2}] as a function of the number of added states $n$, for mixed coherent states $\hat{\rho}_{\text{M}}(n)$ [see Eq.~\eqref{eqn:mixedCSQP}] which saturate the atomic plane $(Q,P)$ in the chaotic-energy region $\epsilon_{\text{M}}=1$. Vertical dashed black lines indicate the value of $n$ where the Husimi projections are shown. The system size is $j=30$.
}
\label{fig05}
\end{figure*}

%%%%%%%%%%%%%%%%%%%%%%%%%%%%%%%%%%%%%%%%%%%%%%%%%%%%%

\subsection{Localization of Evolved Coherent States}
\label{subsec:LocEvolvedCohStates}

Consider initial Glauber-Bloch coherent states $\hat{\rho}_{\bm{x}}=\dyad{\bm{x}}$, 
with coordinates $\bm{x}=(q_{0},p_{0};Q_{0},P_{0})$ and mean energy $\epsilon_{\bm{x}}=h_\text{cl}(\bm{x})$, which are highly localized in the phase space. The time evolution of $\hat{\rho}_{\bm{x}}$ is given by
\begin{equation}
\label{eqn:pureCS}
\hat{\rho}_{\bm{x}}(t)=\hat{U}(t)\dyad{\bm{x}}\hat{U}^{\dagger}(t),
\end{equation}
where $\hat{U}^{\dagger}(t)=e^{-i\hat{H}_D t}$. We study how the initial state $\hat{\rho}_{\bm{x}}$ delocalizes as it evolves in time by considering the R\'enyi occupations $\mathfrak{L}_{2}(\mathcal{A},\hat{\rho}_{\bm{x}}(t))$ and $\mathfrak{L}_2(\epsilon_{\bm{x}},\hat{\rho}_{\bm{x}}(t))$. In Fig.~\ref{fig03}~(a) we see that $\mathfrak{L}_{2}(\mathcal{A},\hat{\rho}_{\bm{x}}(t))$ quickly saturates to 1, indicating that the evolved coherent state becomes fully delocalized in the atomic subspace $\mathcal{A}$. On the other hand, the measure $\mathfrak{L}_2(\epsilon_{\bm x},\hat{\rho}_{\bm{x}}(t))$ 
saturates to a value of $1/2$, indicating that an evolved coherent state never delocalizes completely, being able to cover at most half of the energy shell at large times. This is in accord with the results of the Ref.~\cite{Pilatowsky2021NatCommun}, where it was found that $\mathfrak{L}_2\lesssim 1/2 $ for any pure state, and  complete delocalization  within a classical energy shell can only  be reached   when temporal averages are performed. 

For this reason, in Fig.~\ref{fig03}~(b), we plot the values of the R\'enyi occupations $\mathfrak{L}_{2}(\mathcal{A},\overline{\rho}_{\bm{x}}(T)))$ and $\mathfrak{L}_2(\epsilon_{\bm{x}},\overline{\rho}_{\bm{x}}(T))$, for the time-averaged state
\begin{equation}
\label{eqn:mixedCS}
\overline{{\rho}}_{\bm x}(T)=\frac{1}{T} \int_0^T\dif t\,\hat{\rho}_{\bm x}(t).
\end{equation}

We see that the R\'enyi occupation $\mathfrak{L}_2(\epsilon_{\bm{x}},\overline{\rho}_{\bm{x}}(T))$ saturates to 1 at a time larger than the saturation time of the instantaneous localization of the evolved state $\hat{\rho}(t)$.  In contrast, for $\mathfrak{L}_{2}(\mathcal{A},\overline{\rho}_{\bm{x}}(T))$, a dip appears around $T=2$, which is absent in $\mathfrak{L}_{2}(\epsilon_{\bm x},\overline{\rho}_{\bm{x}}(T))$. This is explained by the atomic and bosonic projections, shown in the top and middle panels of Fig.~\ref{fig03}~(b), respectively. At $T=1$, the atomic projection of $\overline{\rho}_{\bm{x}}(T)$ looks like a closed orbit, which starts to retrace itself at $T=2$, producing an apparent increased localization which is clearly visible as a bright area in the third atomic projection. However, this apparently closed orbit is not closed in the complete phase space. One sees from the bosonic projection in $T=2$ that, in fact, the state is not revisiting the same region but delocalizing over the bosonic variables. This effect is absent in for $\mathfrak{L}_2(\epsilon_{\bm{x}},\overline{\rho}_{\bm{x}}(T))$, where no projection is performed. In that case, the temporal average only smooths out the quantum fluctuations, allowing for a less abrupt route to saturation with a  slightly larger saturation time.

\subsection{Separation of Coherent States in the Atomic and Bosonic Planes}
\label{subsec:SeparationPlanes}

To further study the different behaviors of the R\'enyi occupations $\mathfrak{L}_{2}(\mathcal{A},\hat{\rho})$ and $\mathfrak{L}_2(\epsilon,\hat{\rho})$, we investigate what happens when we consider a mixed state formed by two  coherent states whose centroids are gradually separated  in different directions. By doing this,  we want to investigate how sensible the measures are  to a gradual delocalization in both  the atomic $(Q,P)$ and bosonic $(q,p)$ coordinate planes. We consider  the following mixed state of a pair of coherent states
\begin{equation}
\label{eqn:mixedCSD}
\hat{\rho}_{\text{M}}(D)=\frac{1}{2}(\hat{\rho}_{\bm x}+\hat{\rho}_{\bm y}),
\end{equation}
where the state $\hat{\rho}_{\bm x}=\dyad{\bm x}$ remains fixed at position $\bm x$, while the position $\bm y$ of state $\hat{\rho}_{\bm y}=\dyad{\bm y}$ is varied, and  their phase-space distance given by
\begin{equation}
D=\sqrt{(q_{\bm x}-q_{\bm y})^{2}+(p_{\bm x}-p_{\bm y})^{2}+\Theta^{2}},
\end{equation}
where
\begin{equation}
\cos(\Theta)=\cos(\theta_{\bm x})\cos(\theta_{\bm y})+\cos(\phi_{\bm x}-\phi_{\bm y})\sin(\theta_{\bm x})\sin(\theta_{\bm y}),
\end{equation}
with $\cos(\theta)=1-(Q^{2}+P^{2})/2$ and $\tan(\phi)=-P/Q$. The coordinates of $\bm y$ start with both states at the same position $D=0$, and are changed such that the mean energy $\epsilon_{\text{M}}$ and energy width of $|\bm y\rangle$ remain constant. These conditions ensure that the changes in the occupation measures come exclusively  from the separation of the coherent state and not from changes in the properties of the coherent states $|\bm y\rangle$. 

First we separate the states in the atomic plane $(Q,P)$ by   letting $p_{\bm y}=p_{\bm x}$ be constant, and changing $q_{\bm y}$ slightly to maintain the mean energy of $|\bm y\rangle$ constant. Figure~\ref{fig04}~(a) shows the projection of $\hat{\rho}_{\text{M}}(D)$ in both the atomic and bosonic coordinate planes (top squared panels), and the values of the R\'enyi occupations  as a function of the separation $D$ (bottom rectangular panel). The R\'enyi occupations are divided by their values at $D=0$, $\mathfrak{L}_{2}(\mathcal{A},\hat{\rho}_{\text{M}})/\mathfrak{L}_2(\mathcal{A},\hat{\rho}_{\text{M}}(0))$ and $\mathfrak{L}_2(\epsilon_{\text{M}},\hat{\rho}_{\text{M}})/\mathfrak{L}_2(\epsilon_{\text{M}},\hat{\rho}_{\text{M}}(0))$, allowing to compare their relative growth.  We see that in this case both R\'enyi occupations behaves similarly, increasing to twice their original value.

Now we separate the states in the bosonic plane $(q,p)$, letting  $Q_{\bm y}=Q_{\bm x}$ and $P_{\bm y}=P_{\bm x}$ remain constant, as shown in Fig.~\ref{fig04}~(b). Contrary  to the previous case, the results  show  remarkable differences between $\mathfrak{L}_2(\mathcal{A},\hat{\rho}_{\text{M}})$ and  $\mathfrak{L}_{2}(\epsilon_{\text{M}},\hat{\rho}_{\text{M}})$. While the R\'enyi occupation $\mathfrak{L}_2(\epsilon_{\text{M}},\hat{\rho}_{\text{M}})$ behaves as in the previous case, doubling its value, $\mathfrak{L}_{2}(\mathcal{A},\hat{\rho}_{\text{M}})$ remains constant. We see that, the R\'enyi occupation $\mathfrak{L}_{2}(\mathcal{A},\hat{\rho}_{\text{M}})$ is not sensitive  to delocalization in the bosonic plane, which is a consequence of the partial integration over the bosonic variables $(q,p)$ in Eq.~\eqref{eqn:Husimiprojection} that erases the information about delocalization in the bosonic plane.

\subsection{Saturation of Atomic Plane}
\label{subsec:SatAtomicPlane}

We now focus on how $\mathfrak{L}_{2}(\mathcal{A},\hat{\rho})$ and $\mathfrak{L}_2(\epsilon,\hat{\rho})$ respond to a decreasing localization in the atomic plane $(Q,P)$ but letting the state well localized in the bosonic one $(q,p)$. We consider a mixed state comprised of $n$ coherent states  
\begin{equation}
\label{eqn:mixedCSQP}
\hat{\rho}_{\text{M}}(n)=\frac{1}{n}\sum_{i=1}^{n}\dyad{\bm{ x}_i},
\end{equation}
whose centroids $\bm{x}_i$ are homogeneously distributed over an increasing area of  the Bloch sphere, as shown in the top panels of Fig.~\ref{fig05}, but have the same classical energy $\epsilon_{\text{M}}= h_\text{cl}(\bm x_i)$.

We computed $\mathfrak{L}_{2}(\mathcal{A},\hat{\rho}_{\text{M}})$ and $\mathfrak{L}_2(\epsilon_{\text{M}},\hat{\rho}_{\text{M}})$ as we increase the Bloch sphere area occupied by the mixed state by increasing $n$ and plot the result in Fig.~\ref{fig05}. Note that the R\'enyi occupation $\mathfrak{L}_{2}(\mathcal{A},\hat{\rho}_{\text{M}})$ saturates completely to unity when we fill completely the Bloch sphere, even if  the states are highly localized in the bosonic plane. On the other hand, the R\'enyi occupation $\mathfrak{L}_2(\epsilon_{\text{M}},\hat{\rho}_{\text{M}})$ is sensitive to the localization of the mixed state in the bosonic plane, and consequently it reaches a very small  value of $\mathfrak{L}_{2}\sim0.07$, indicating that the state $\hat{\rho}_{\text{M}}$ only covers $\sim7\%$ of the  phase-space volume of the energy shell at $\epsilon_{\text{M}}$. This is understandable when we see the Husimi projections. Even when the atomic projection fills completely the Bloch sphere, the bosonic projections show  horizontal tubular shapes which  cover the available bosonic space only partially.

These results confirm that the Husimi projection over the atomic subspace used in the R\'enyi occupation $ \mathfrak{L}_{2}(\mathcal{A},\hat{\rho}_{\text{M}})$ loses possible localization of the states in the bosonic plane, which is not the case for the  R\'enyi occupation $\mathfrak{L}_2(\epsilon_{\text{M}},\hat{\rho}_{\text{M}})$, which is sensitive to the localization of the states in both planes. In summary, for states well localized in energy, such as eigenstates and coherent sates considered here, $\mathfrak{L}_2(\epsilon,\hat{\rho})$ is a more sensitive measure. However, for states that are delocalized in energy, $\mathfrak{L}_2(\epsilon,\hat{\rho})$ could lose important information about the localization of the states in energy, and $ \mathfrak{L}_{2}(\mathcal{A},\hat{\rho})$ may be more adequate.

%%%%%%%%%%%%%% CONCLUSIONS %%%%%%%%%%%%%%%%

\section{CONCLUSIONS}
\label{sec:Conclusions}

In this paper we have put forward a general mathematical framework to define  occupation (and from it localization) of quantum states with respect to a given discrete or continuous measure space. Since the measures obtained are related to the R\'enyi entropies of order $\alpha \geq 0$,  we called them  R\'enyi occupations  of order $\alpha$. We showed that  this mathematical framework includes two recently employed localization measures in phase space, used to gauge localization of eigenstates and coherent states in the chaotic region of the Dicke model. One of them measures localization of the Husimi function projected into the atomic phase space, whereas the other measures the localization of the Husimi function over the energy shells of the corresponding classical model.

Both R\'enyi occupations were compared in detail and the origin of their  differences was elucidated. It was shown that the occupation over the atomic subspace can miss information about the localization of states in the bosonic plane. For states well localized in energy, such as eigenstates and coherent states, the occupation over classical energy shells is a more suitable measure of localization in phase space.

We emphasize, however, that there is no a unique way to define localization in phase space. In this article, we focus on the R\'enyi occupations of order  $\alpha=2$, but orders $\alpha\neq2$ can be used to reveal other aspects of the Husimi distribution either in classical energy shells or in its projection in the atomic subspace. In general, R\'enyi occupations with $\alpha>1$ strengthen the contribution of regions with largest distribution values in detriment of those with smaller ones; and conversely, R\'enyi occupations of order $\alpha<1$ tend to equally weight the contributions of the regions where the probability distribution is different from zero, independently of its value. A study of the dependence of the R\'enyi occupations on $\alpha$ would allow to reveal multifractality in the occupation of the states in phase space.

Another interesting issue to be explored in the future is the relationship between the phase-space localization of non-stationary states, as revealed by the R\'enyi occupations introduced here, and the breaking of the eigenstate thermalization hypothesis (ETH)~\cite{Nandkishore2015,Abanin2019}. It is conjectured that a necessary condition for the ETH to be valid is that the infinite-time average of generic non-stationary states reaches an energy-shell R\'enyi occupation equal to one. How can the localization measure help to identify states that violate the ETH is an open question to be explored in future works.

%%%%%%%%%%%%%% ACKNOWLEDGMENTS %%%%%%%%%%%%%%%%

\section*{ACKNOWLEDGMENTS}

We thank Lea F. Santos for her insightful comments. We acknowledge the support of the Computation Center - ICN, in particular to Enrique Palacios, Luciano D\'iaz, and Eduardo Murrieta. DV, SP-C, and JGH acknowledge financial support from the DGAPA- UNAM project IN104020, and SL-H from the Mexican CONACyT project CB2015-01/255702.

%%%%%%%%%%%%%% APPENDIX. %%%%%%%%%%%%%%%%

\appendix

\section{Derivation of the R\'enyi Volume}
\label{app:LocalizationMeasuresDetails}

As in Sec.~\ref{sec:LocalizationMeasures}, we consider a space $X$ with a measure $\mathcal{V}$, which generates a (Lebesgue) integral $\int_X \dif \mathcal{V}(\bm x) \,\bullet$. We will construct the R\'enyi volume $\mathcal{V}_\alpha(X,\varphi)$ given by Eqs.~\eqref{eqn:OcuppiedVolumeOrder1}~and~\eqref{eqn:OcuppiedVolumeOrderalpha} by requiring the function $\mathcal{V}(X,\varphi)$ to be equal to  $\mathcal{V}(\Omega)$  for a uniform distribution over $\Omega\subseteq X$, be bounded by the volume of $X$, and be homogeneous under the scaling of $\mathcal{V}$.  

The easiest function that satisfies these properties is $\mathcal{V}_0(X,\varphi)=\mathcal{V}(\text{supp}(\varphi))$, where  $\text{supp}(\varphi)=\{\bm x\in X\mid \varphi(\bm x)\neq 0\}$ [which is the limit $\alpha\to 0$ of Eq.~\eqref{eqn:OcuppiedVolumeOrderalpha}]. However,  this may not be very useful, because $\varphi$ may be very small --but not zero-- in some region of $X$. We would want $\mathcal{V}(X,\varphi)$ to be able to ignore those low-probability regions.

The threshold for when to consider a probability {\it ignorable} is, of course, arbitrary. Let us consider any function $f\colon(0,\infty)\to \mathbb{R}$, so that the value $f(x)$ establishes how much we care about the regions where $\varphi(\bm x)$ is less than $x$. Then, it is natural to define the weighted mean
\begin{equation}
F(\varphi)=\int_X \dif \mathcal{V}(\bm x) \varphi(\bm x)f(\varphi(\bm x)).
\end{equation}

For a bounded measurable subset $\Omega\subseteq X$, define the uniform probability distribution $\varphi_\Omega(\bm x)$  as in Eq.~\eqref{eq:unif_prob}. For these uniform probabilities, we require $\mathcal{V}_f(X,\varphi_\Omega)=\mathcal{V}(\Omega)$. This condition will allow us to find $\mathcal{V}_f(X,\varphi)$ for any normalized distribution $\varphi$. 

We have
\begin{equation}
F(\varphi_\Omega) =\int_\Omega \dif \mathcal{V}(\bm x)\varphi_\Omega(\bm x) f(\varphi_\Omega(\bm x))=f\left(\frac{1}{\mathcal{V}_f(X,\varphi_\Omega)}\right).
\end{equation}
If we require $f$ to be invertible, then $\mathcal{V}_f(X,\varphi_\Omega)=1/f^{-1}(F(\varphi_\Omega)).$ Thus, it is natural to extend this formula and define
\begin{align}
    \label{eq:general_L_definition}
    \mathcal{V}_f(X,\varphi)&=\frac{1}{f^{-1}(F(\varphi))}\\&=\left[f^{-1}\bigg(\int_X \dif \mathcal{V}(\bm x)\varphi(\bm x) f(\varphi(\bm x))\bigg)\right]^{-1}\!\!\! \nonumber,
\end{align}
for any distribution $\varphi$. Note that because $f^{-1}>0$, we have $\mathcal{V}_f(X,\varphi)>0$. However, without imposing some extra conditions on $f$, we cannot say anything about the upper bound of $\mathcal{V}_f(X,\varphi)$.

Because we wish to interpret $\mathcal{V}_f(X,\varphi)$ as the volume of the region occupied by $\varphi$, so we need to first ensure that $\mathcal{V}_f(X,\varphi)\leq \mathcal{V}(X)$ when $X$ is bounded.
To guarantee this, we require that 
\makeatletter
\newcommand{\myitem}[1]{%
\item[#1]\protected@edef\@currentlabel{#1}%
}
\makeatother
\begin{enumerate}
    \myitem{($\star$)} \label{condition:f_increasing}$f(x)$ is strictly increasing and $g(x)=xf(x)$ is convex. 
\end{enumerate}

Then, by Jensen's inequality (c.f.~\cite{Durrett2019}, p. 21), the convexity of $g(x)=xf(x)$ guarantees that
\begin{align}
\label{eq:Jenseninequalityapplication}
F(\varphi)&=\int_X \dif \mathcal{V}(\bm x) g(\varphi(\bm x)) \\
&\geq \mathcal{V}(X)\,g\bigg(\frac{1}{\mathcal{V}(X)}\int_X \dif \mathcal{V}(\bm x) \varphi(\bm x)\bigg)=f\bigg(\frac{1}{\mathcal{V}(X)}\bigg)\nonumber
\end{align}
for any distribution $\varphi$ with $\int_X \dif \mathcal{V}(\bm x) \varphi(\bm x)=1$. Because $f^{-1}$ is increasing if $f$ is increasing, we get
\begin{align}
    \mathcal{V}_f(X,\varphi)&=\frac{1}{f^{-1}\big(F(\varphi)\big)}\leq \frac{1}{f^{-1}\Big(f\big(1/\mathcal{V}(X)\big)\Big)}=\mathcal{V}(X)\nonumber \\
    &=\mathcal{V}_f(X,\varphi_X),
\end{align}
where $\varphi_X$ is the uniform distribution on $X$.

Note that the argument above also works if
\begin{enumerate}
    \myitem{($\star\star$)} \label{condition:f_decreasing}$f(x)$ is strictly decreasing and $g(x)=xf(x)$ is concave
\end{enumerate}
 by replacing $f\to -f$.

Moreover, Jensen's inequality in Eq.~\eqref{eq:Jenseninequalityapplication}  becomes an equality if and only if there exist $a,b\in \mathbb{R}$ such that $xf(x)=g(x)=a+bx$ for all $x\in \{\varphi(\bm x) \mid \bm x\in X\}$ (i.e. $g$ is linear in the range of $\varphi$). If any of the stricter conditions
\begin{enumerate}[label=(A\arabic*]
    \myitem{($\star'$)} \label{condition:f_increasing_strict}$f(x)$ is strictly increasing and $g(x)=xf(x)$ is strictly convex, or
    \myitem{($\star\star'$)} \label{condition:f_decreasing_strict}$f(x)$ is strictly decreasing and $g(x)=xf(x)$ is strictly concave,
\end{enumerate}
is satisfied, then $g$ cannot be linear in $\{\varphi(\bm x) \mid \bm x\in X\}$ unless that set consists of a single point, in which case $\varphi$ must be constant. So any of conditions~\ref{condition:f_increasing_strict}~or~\ref{condition:f_decreasing_strict} implies that $\mathcal{V}_f(X,\varphi)=\mathcal{V}(X)$ occurs only when $\varphi=\varphi_X$ is the uniform distribution. 
 
If $f_1(x)=\log(x)$, condition~\ref{condition:f_increasing_strict} is satisfied and we obtain Eq.~\eqref{eqn:OcuppiedVolumeOrder1}. For  $f_\alpha(x)=x^{\alpha-1}$ with $\alpha\geq 0$ and $\alpha\neq 1$, condition~\ref{condition:f_increasing_strict} is satisfied if $\alpha> 1$, condition~\ref{condition:f_decreasing_strict} is satisfied if $0<\alpha< 1$, and  condition~\ref{condition:f_decreasing} [but not condition~\ref{condition:f_decreasing_strict}] is satisfied if $\alpha = 0$. In any case, Eq.~\eqref{eq:general_L_definition} yields Eq.~\eqref{eqn:OcuppiedVolumeOrderalpha}.
  
There are many other functions $f$ besides $f_1(x)=\log(x)$ and $f_\alpha(x)=x^{\alpha-1}$ ($\alpha> 0$, $\alpha\neq 1$) that satisfy the conditions~\ref{condition:f_increasing_strict}~or~\ref{condition:f_decreasing_strict}. However, the measures  $\mathcal{V}_\alpha(X,\varphi)=\mathcal{V}_{f_\alpha}(X,\varphi)$ given by Eqs.~\eqref{eqn:OcuppiedVolumeOrderalpha}~and~\eqref{eqn:OcuppiedVolumeOrder1} are the only ones that are homogeneous under scaling of the measure $\mathcal{V}$ [see Eq.~\eqref{eq:scaling_property}]. If $\mathcal{V}_f(X,\varphi)$ satisfies Eq.~\eqref{eq:scaling_property}, it can be shown that $f=a f_\alpha +b$ where $a,b\in \mathbb{R}$ with $a\neq 0$ --see Theorem (5.2.19) and Eq.~(5.2.26) of Ref.~\cite{Aczel1975}--. It is  straightforward to verify that $f'=a f +b$  implies $\mathcal{V}_{f'}(X,\varphi)=\mathcal{V}_f(X,\varphi)$ directly from Eq.~\eqref{eq:general_L_definition}. In fact, the converse is also true --see  Theorem (5.2.16) of Ref.~\cite{Aczel1975}--. Thus, if $\mathcal{V}_f(X,\varphi)$ satisfies Eq.~\eqref{eq:scaling_property}, then $\mathcal{V}_{f}(X,\varphi)=\mathcal{V}_\alpha(X,\varphi)$ for some $\alpha\geq 0$.

\section{Diagonalization basis}
\label{app:DiagonalizationBasis}

\subsection{Fock Basis}

The standard Fock basis is given by the tensor product $|n\rangle\otimes|j,m_{z}\rangle$, where $|n\rangle\,$ ($n=0,1,2,\ldots$) are the eigenstates of the number operator $\hat{n}=\hat{a}^{\dagger}\hat{a}$ of the infinite dimensional bosonic subspace, and $|j,m_{z}\rangle\,$ ($m_{z}=-j,-j+1,\ldots,j-1,j$) are the eigenstates of the collective pseudo-spin operator $\hat{J}_{z}$ of the finite dimensional pseudo-spin subspace with Hilbert-space dimension $2j+1$. As the bosonic subspace has infinite Hilbert-space dimension, we need to truncate it by choosing an maximal excitation level $n_{\text{max}}$, which enables us to define a finite dimension $n_{\text{max}}+1$. Then, the global Hilbert-space dimension is given by $\mathcal{D}_{\text{FB}}=(n_{\text{max}}+1)\times(2j+1)$. 

For the  system size $j=30$ selected in this work, we required $n_{\text{max}}=420$ with global dimension $\mathcal{D}_{\text{FB}}=25681$ to ensure $\mathcal{D}_{\text{FB}}^{c}=8030$ converged eigenstates and eigenvalues. For the selected Hamiltonian parameters, the energy spectrum ranges from the ground-state energy $\epsilon_\text{GS}=-2.125$ up to a truncated converged energy $\epsilon_\text{T}=3.683$.

An advantage of using the Fock basis is that one may compute exact projections of the Husimi function of a state $\hat{\rho}$ in both atomic and bosonic subspaces of the Dicke model by using the closed expressions~\cite{Deaguiar1991,Furuya1992}
\begin{equation}
\begin{split}
\widetilde{\mathcal{Q}}_{\hat{\rho}}(Q,P) = & \frac{j}{2\pi} \int\int \dif q \dif p \mathcal{Q}_{\hat{\rho}}(q,p;Q,P) \\
= & A(Q,P) \sum_{n=0}^{n_{\text{max}}} \sum_{m_{z}=-j}^{j} \sum_{m'_{z}=-j}^{j} \left\{ (c_{n,m_{z}}^{\hat{\rho}})^{\ast} c_{n,m'_{z}}^{\hat{\rho}} \times \right. \\  & \left. G_{m_{z}}^{A}(Q,P) (G_{m'_{z}}^{A}(Q,P))^{\ast} \right\},
\end{split}
\end{equation}
with $A(Q,P)=\left(1-\frac{Q^{2}+P^{2}}{4}\right)^{2j}$, and
\begin{equation}
G_{m_{z}}^{A}(Q,P) = \sqrt{\left(\begin{array}{c} 2j \\ j + m_{z} \end{array}\right)} \left(\frac{Q+iP}{\sqrt{4-Q^{2}-P^{2}}}\right)^{j+m_{z}},
\end{equation}
for the atomic projection, and
\begin{equation}
\begin{split}
\widetilde{\mathcal{Q}}_{\hat{\rho}}(q,p) = & \frac{2j+1}{4\pi} \int\int \dif Q \dif P \mathcal{Q}_{\hat{\rho}}(q,p;Q,P) \\
= & B(q,p) \sum_{n=0}^{n_{\text{max}}} \sum_{n'=0}^{n_{\text{max}}} \sum_{m_{z}=-j}^{j} \left\{ (c_{n,m_{z}}^{\hat{\rho}})^{\ast} c_{n',m_{z}}^{\hat{\rho}} \times \right. \\  & \left. G_{n}^{B}(q,p) (G_{n'}^{B}(q,p))^{\ast} \right\},
\end{split}
\end{equation}
with $B(q,p)=\text{exp}\left(-\frac{j}{2}(q^{2}+p^{2})\right)$, and
\begin{equation}
G_{n}^{B}(q,p) = \frac{1}{\sqrt{n!}} \left(\sqrt{\frac{j}{2}}(q+ip)\right)^{n}
\end{equation}
for the bosonic one. In both expressions, $c_{n,m_{z}}^{\hat{\rho}}$ are the coefficients of the state $\hat{\rho}$ expanded in this basis.

\subsection{Efficient Basis}

The efficient basis is obtained by taking the eigenbasis of the Dicke Hamiltonian $\hat{H}_{D}$ [see Eq.~\eqref{eqn:qua_hamiltonian}] in the limit $\omega_{0}\rightarrow0$. The first step to construct this basis is to define a displaced annihilation operator  $\hat{A}=\hat{a}+(2\gamma/(\omega\sqrt{\mathcal{N}}))\hat{J}_{x}$. Then, a rotation in the Bloch sphere $(\hat{J}_{x},\hat{J}_{y},\hat{J}_{z})\rightarrow(\hat{J}'_{z},\hat{J}'_{y},-\hat{J}'_{x})$ is performed. Finally, taking the limit $\omega_{0}\rightarrow0$, one obtains a basis given by the tensor product $|N\rangle\otimes|j,m_x\rangle$, where the states are explicitly 
\begin{equation}
|N\rangle \otimes |j,m_x\rangle = \frac{(\hat{A}^{\dagger})^{N}}{\sqrt{N!}} |N=0\rangle \otimes |j,m_x\rangle,     
\end{equation}
with $|N\rangle$ the eigenstates of the operator $\hat{A}^{\dagger}\hat{A}$, $|N=0\rangle \otimes |j,m_x\rangle$ the vacuum state of the modified bosonic subspace which defines coherent states $|N=0\rangle=|-2\gamma m_x/(\omega\sqrt{\mathcal{N}})\rangle$ in the standard Fock basis for each value of $m_x$, which corresponds to the rotated eigenvalue of the original collective pseudo-spin operator $\hat{J}_{x}$.

The states of the pseudo-spin subspace are defined in the same way as in the Fock basis $|j,m_x\rangle\,$ ($m_x=-j,-j+1,\ldots,j-1,j$) with Hilbert-space dimension $2j+1$. As in the Fock basis, there are infinitely many states corresponding to the modified bosonic subspace $|N\rangle\,$  ($N=0,1,2,\ldots$), so a truncation is also necessary. By choosing a maximal excitation level $N_{\text{max}}$, we determine a finite dimension $N_{\text{max}}+1$, so that the global Hilbert-space dimension for this basis is given by $\mathcal{D}_{\text{EB}}=(N_{\text{max}}+1)\times(2j+1)$.

For the selected system size $j=30$, we required $N_{\text{max}}=200$ with global dimension $\mathcal{D}_{\text{EB}}=12261$, to ensure $\mathcal{D}_{\text{EB}}^{c}=8041$ converged eigenstates and eigenvalues. For these Hamiltonian parameters, the energy spectrum ranges from the ground-state energy $\epsilon_\text{GS}=-2.125$ up to a truncated converged energy $\epsilon_\text{T}=3.688$.

Although not used in this work, it is worth noting that the efficient basis allows to reach  larger system sizes ($j\sim100$) in the superradiant phase of the model, which would be unreachable by using the standard Fock basis~\cite{Chen2008,Bastarrachea2014a,Bastarrachea2014b}.

\section{Lower Bound on Phase-Space Localization}
\label{app:LowerBoundOnPhaseSpaceLocalization}

The Lieb conjecture~\cite{Lieb1978,Lieb2014} guarantees that the coherent states $\hat{\rho}_{\bm x}=\dyad{\bm x}$ are the most localized in the phase space of the Dicke model. Thus, $\mathcal{V}_\alpha(\mathcal{M},\hat{\rho}_{\bm x})$ is a lower bound on the R\'enyi volume in phase space for any state. We will this value as given by Eqs.~\eqref{eqn:OcuppiedVolumeOrderalpha}~and~\eqref{eqn:OcuppiedVolumeOrder1}. Because all coherent states are translations of each other, and $\mathcal{V}$ is invariant under translations in $\mathcal{M}$ ($\mathcal{V}$ is the Haar measure for the group of translations of $\mathcal{M}$), it suffices to calculate $\mathcal{V}_\alpha(\mathcal{M},\hat{\rho}_{\bm c})$ for one coherent state $\hat{\rho}_{\bm c}=\dyad{\bm c}$ centered at $\bm c\in\mathcal{M}$. For convenience, let us choose $\bm c=(0,0,0,0)$, so $\ket{\bm c}=\ket{q=p=Q=P=0}$. The Husimi function of $\ket{\bm c}$ is given by~\cite{Arecchi1972,Villasenor2020}
\begin{equation}
\label{eqn:HusimiOfCoherentStateAtOrigin}
    \mathcal{Q}_{\bm c}(\bm x)=\exp(-\frac{j}{2}\left(q^2+p^2\right))\bigg(1-\frac{P^2+Q^2}{4}\bigg)^{2j}.
\end{equation}

Inserting Eq.~\eqref{eqn:HusimiOfCoherentStateAtOrigin} into  Eq.~\eqref{eqn:OcuppiedVolumeOrderalpha} and performing the integration, one gets
\begin{align}
    \mathcal{V}_\alpha(\mathcal{M},\hat{\rho}_{\bm c})&=8 \pi ^2 \hbar _{\text{eff}}^2 \left(\hbar _{\text{eff}}+2\right){}^{\frac{\alpha }{1-\alpha }} \left(\alpha  \left(2 \alpha +\hbar \nonumber _{\text{eff}}\right)\right){}^{\frac{1}{\alpha -1}}\\
    &=(2 \pi  \hbar_\text{eff})^2 \alpha^\frac{2}{\alpha-1}  + O(\hbar_\text{eff}^3), 
\end{align}
where $\hbar_\text{eff}=1/j$. Using $\lim_{\alpha\to 1}  \alpha ^{\frac{2 }{\alpha -1}}=e^2$, we get $\mathcal{V}_1(\mathcal{M},\hat{\rho}_{\bm c})=(2 \pi  \hbar_\text{eff}\,e)^2 + O(\hbar_\text{eff}^3) \nonumber$, which can also be verified by directly integrating Eq.~\eqref{eqn:OcuppiedVolumeOrder1},
\begin{align}
\mathcal{V}_1(\mathcal{M},\hat{\rho}_{\bm c})= \frac{(2 \pi \hbar_\text{eff})^2  e^{\frac{\hbar_\text{eff}+4}{\hbar_\text{eff} +2}}}{\hbar_\text{eff}+2}=(2 \pi  \hbar_\text{eff}\,e)^2 + O(\hbar_\text{eff}^3).
\end{align}

The higher order terms $O(\hbar_\text{eff}^3)$ are small provided that ${\alpha \gg \hbar_\text{eff}^2}$, in which case Eq.~\eqref{eq:locMeasureSmallerThath} is valid. If $\alpha$ is of the order of $\hbar_\text{eff}^2$ or smaller, this is not true anymore. In fact, in the limit of $\alpha \to 0$,
\begin{equation}
\mathcal{V}_0(\mathcal{M},\hat{\rho}_{\bm c})=\mathcal{V}(\{\bm x\in \mathcal{M} \mid  \mathcal{Q}_{\bm c}(\bm x)\neq 0\})=\mathcal{V}(\mathcal{M})=\infty.
\end{equation}
This happens because, as $\alpha$ becomes smaller, the R\'enyi volume $\mathcal{V}_\alpha(\mathcal{M},\hat{\rho})$ looses the ability to differentiate between the regions where  $\mathcal{Q}_{\hat{\rho}}$ is big and the regions where it is small but non-zero. Thus, the R\'enyi volumes in  phase space $\mathcal{V}_\alpha(\mathcal{M},\hat{\rho})$ are more useful when $\alpha \gg \hbar_\text{eff}^2$.

%----------------------------------------------------------------------------------------
%	REFERENCE LIST
%----------------------------------------------------------------------------------------
%%%%%%%%%%%%%%%%%%%%%%%%%%%%%%%%%%%%%%%%%%%%%%%%%%
\bibliography{main}

\begin{thebibliography}{85}%
\makeatletter
\providecommand \@ifxundefined [1]{%
 \@ifx{#1\undefined}
}%
\providecommand \@ifnum [1]{%
 \ifnum #1\expandafter \@firstoftwo
 \else \expandafter \@secondoftwo
 \fi
}%
\providecommand \@ifx [1]{%
 \ifx #1\expandafter \@firstoftwo
 \else \expandafter \@secondoftwo
 \fi
}%
\providecommand \natexlab [1]{#1}%
\providecommand \enquote  [1]{``#1''}%
\providecommand \bibnamefont  [1]{#1}%
\providecommand \bibfnamefont [1]{#1}%
\providecommand \citenamefont [1]{#1}%
\providecommand \href@noop [0]{\@secondoftwo}%
\providecommand \href [0]{\begingroup \@sanitize@url \@href}%
\providecommand \@href[1]{\@@startlink{#1}\@@href}%
\providecommand \@@href[1]{\endgroup#1\@@endlink}%
\providecommand \@sanitize@url [0]{\catcode `\\12\catcode `\$12\catcode
  `\&12\catcode `\#12\catcode `\^12\catcode `\_12\catcode `\%12\relax}%
\providecommand \@@startlink[1]{}%
\providecommand \@@endlink[0]{}%
\providecommand \url  [0]{\begingroup\@sanitize@url \@url }%
\providecommand \@url [1]{\endgroup\@href {#1}{\urlprefix }}%
\providecommand \urlprefix  [0]{URL }%
\providecommand \Eprint [0]{\href }%
\providecommand \doibase [0]{http://dx.doi.org/}%
\providecommand \selectlanguage [0]{\@gobble}%
\providecommand \bibinfo  [0]{\@secondoftwo}%
\providecommand \bibfield  [0]{\@secondoftwo}%
\providecommand \translation [1]{[#1]}%
\providecommand \BibitemOpen [0]{}%
\providecommand \bibitemStop [0]{}%
\providecommand \bibitemNoStop [0]{.\EOS\space}%
\providecommand \EOS [0]{\spacefactor3000\relax}%
\providecommand \BibitemShut  [1]{\csname bibitem#1\endcsname}%
\let\auto@bib@innerbib\@empty
%</preamble>
\bibitem [{\citenamefont {Chirikov}\ \emph {et~al.}(1981)\citenamefont
  {Chirikov}, \citenamefont {Izrailev},\ and\ \citenamefont
  {Shepelyansky}}]{Chirikov1981}%
  \BibitemOpen
  \bibfield  {author} {\bibinfo {author} {\bibfnamefont {B.V.}\ \bibnamefont
  {Chirikov}}, \bibinfo {author} {\bibfnamefont {F.~M.}\ \bibnamefont
  {Izrailev}}, \ and\ \bibinfo {author} {\bibfnamefont {D.~L.}\ \bibnamefont
  {Shepelyansky}},\ }\bibfield  {title} {\enquote {\bibinfo {title} {Dynamical
  stochasticity in classical and quantum mechanics},}\ }\href@noop {}
  {\bibfield  {journal} {\bibinfo  {journal} {Sov. Scient. Rev. C}\ }\textbf
  {\bibinfo {volume} {2}},\ \bibinfo {pages} {209--267} (\bibinfo {year}
  {1981})}\BibitemShut {NoStop}%
\bibitem [{\citenamefont {Izrailev}(1990)}]{Izrailev1990}%
  \BibitemOpen
  \bibfield  {author} {\bibinfo {author} {\bibfnamefont {F.~M.}\ \bibnamefont
  {Izrailev}},\ }\bibfield  {title} {\enquote {\bibinfo {title} {Simple models
  of quantum chaos: Spectrum and eigenfunctions},}\ }\href {\doibase
  10.1016/0370-1573(90)90067-C} {\bibfield  {journal} {\bibinfo  {journal}
  {Phys. Rep.}\ }\textbf {\bibinfo {volume} {196}},\ \bibinfo {pages}
  {299--392} (\bibinfo {year} {1990})}\BibitemShut {NoStop}%
\bibitem [{\citenamefont {Casati}\ \emph {et~al.}(1984)\citenamefont {Casati},
  \citenamefont {Chirikov},\ and\ \citenamefont {Shepelyansky}}]{Casati1984a}%
  \BibitemOpen
  \bibfield  {author} {\bibinfo {author} {\bibfnamefont {Giulio}\ \bibnamefont
  {Casati}}, \bibinfo {author} {\bibfnamefont {B.~V.}\ \bibnamefont
  {Chirikov}}, \ and\ \bibinfo {author} {\bibfnamefont {D.~L.}\ \bibnamefont
  {Shepelyansky}},\ }\bibfield  {title} {\enquote {\bibinfo {title} {Quantum
  limitations for chaotic excitation of the hydrogen atom in a monochromatic
  field},}\ }\href {\doibase 10.1103/PhysRevLett.53.2525} {\bibfield  {journal}
  {\bibinfo  {journal} {Phys. Rev. Lett.}\ }\textbf {\bibinfo {volume} {53}},\
  \bibinfo {pages} {2525--2528} (\bibinfo {year} {1984})}\BibitemShut {NoStop}%
\bibitem [{\citenamefont {Casati}\ \emph {et~al.}(1987)\citenamefont {Casati},
  \citenamefont {Chirikov}, \citenamefont {Shepelyansky},\ and\ \citenamefont
  {Guarnesi}}]{Casati1987}%
  \BibitemOpen
  \bibfield  {author} {\bibinfo {author} {\bibfnamefont {G.}~\bibnamefont
  {Casati}}, \bibinfo {author} {\bibfnamefont {B.V.}\ \bibnamefont {Chirikov}},
  \bibinfo {author} {\bibfnamefont {D.~L.}\ \bibnamefont {Shepelyansky}}, \
  and\ \bibinfo {author} {\bibfnamefont {I.}~\bibnamefont {Guarnesi}},\
  }\bibfield  {title} {\enquote {\bibinfo {title} {Relevance of classical chaos
  in quantum mechanics: The hydrogen atom in a monochromatic field},}\ }\href
  {\doibase 10.1016/0370-1573(87)90009-3} {\bibfield  {journal} {\bibinfo
  {journal} {Phys. Rep.}\ }\textbf {\bibinfo {volume} {154}},\ \bibinfo {pages}
  {77--123} (\bibinfo {year} {1987})}\BibitemShut {NoStop}%
\bibitem [{\citenamefont {Bl\"umel}\ and\ \citenamefont
  {Smilansky}(1987)}]{Blumel1987}%
  \BibitemOpen
  \bibfield  {author} {\bibinfo {author} {\bibfnamefont {R.}~\bibnamefont
  {Bl\"umel}}\ and\ \bibinfo {author} {\bibfnamefont {U.}~\bibnamefont
  {Smilansky}},\ }\bibfield  {title} {\enquote {\bibinfo {title} {Localization
  of floquet states in the rf excitation of {R}ydberg atoms},}\ }\href
  {\doibase 10.1103/PhysRevLett.58.2531} {\bibfield  {journal} {\bibinfo
  {journal} {Phys. Rev. Lett.}\ }\textbf {\bibinfo {volume} {58}},\ \bibinfo
  {pages} {2531--2534} (\bibinfo {year} {1987})}\BibitemShut {NoStop}%
\bibitem [{\citenamefont {Anderson}(1958)}]{Anderson1958}%
  \BibitemOpen
  \bibfield  {author} {\bibinfo {author} {\bibfnamefont {P.~W.}\ \bibnamefont
  {Anderson}},\ }\bibfield  {title} {\enquote {\bibinfo {title} {Absence of
  diffusion in certain random lattices},}\ }\href {\doibase
  10.1103/PhysRev.109.1492} {\bibfield  {journal} {\bibinfo  {journal} {Phys.
  Rev.}\ }\textbf {\bibinfo {volume} {109}},\ \bibinfo {pages} {1492--1505}
  (\bibinfo {year} {1958})}\BibitemShut {NoStop}%
\bibitem [{\citenamefont {Fishman}\ \emph {et~al.}(1982)\citenamefont
  {Fishman}, \citenamefont {Grempel},\ and\ \citenamefont
  {Prange}}]{Fishman1982}%
  \BibitemOpen
  \bibfield  {author} {\bibinfo {author} {\bibfnamefont {Shmuel}\ \bibnamefont
  {Fishman}}, \bibinfo {author} {\bibfnamefont {D.~R.}\ \bibnamefont
  {Grempel}}, \ and\ \bibinfo {author} {\bibfnamefont {R.~E.}\ \bibnamefont
  {Prange}},\ }\bibfield  {title} {\enquote {\bibinfo {title} {Chaos, quantum
  recurrences, and {A}nderson localization},}\ }\href {\doibase
  10.1103/PhysRevLett.49.509} {\bibfield  {journal} {\bibinfo  {journal} {Phys.
  Rev. Lett.}\ }\textbf {\bibinfo {volume} {49}},\ \bibinfo {pages} {509--512}
  (\bibinfo {year} {1982})}\BibitemShut {NoStop}%
\bibitem [{\citenamefont {Casati}\ \emph {et~al.}(1993)\citenamefont {Casati},
  \citenamefont {Chirikov}, \citenamefont {Guarneri},\ and\ \citenamefont
  {Izrailev}}]{Casati1993}%
  \BibitemOpen
  \bibfield  {author} {\bibinfo {author} {\bibfnamefont {G.}~\bibnamefont
  {Casati}}, \bibinfo {author} {\bibfnamefont {B.~V.}\ \bibnamefont
  {Chirikov}}, \bibinfo {author} {\bibfnamefont {I.}~\bibnamefont {Guarneri}},
  \ and\ \bibinfo {author} {\bibfnamefont {F.~M.}\ \bibnamefont {Izrailev}},\
  }\bibfield  {title} {\enquote {\bibinfo {title} {Band-random-matrix model for
  quantum localization in conservative systems},}\ }\href {\doibase
  10.1103/PhysRevE.48.R1613} {\bibfield  {journal} {\bibinfo  {journal} {Phys.
  Rev. E}\ }\textbf {\bibinfo {volume} {48}},\ \bibinfo {pages} {R1613--R1616}
  (\bibinfo {year} {1993})}\BibitemShut {NoStop}%
\bibitem [{\citenamefont {Borgonovi}\ \emph {et~al.}(1996)\citenamefont
  {Borgonovi}, \citenamefont {Casati},\ and\ \citenamefont
  {Li}}]{Borgonovi1996}%
  \BibitemOpen
  \bibfield  {author} {\bibinfo {author} {\bibfnamefont {Fausto}\ \bibnamefont
  {Borgonovi}}, \bibinfo {author} {\bibfnamefont {Giulio}\ \bibnamefont
  {Casati}}, \ and\ \bibinfo {author} {\bibfnamefont {Baowen}\ \bibnamefont
  {Li}},\ }\bibfield  {title} {\enquote {\bibinfo {title} {Diffusion and
  localization in chaotic billiards},}\ }\href {\doibase
  10.1103/PhysRevLett.77.4744} {\bibfield  {journal} {\bibinfo  {journal}
  {Phys. Rev. Lett.}\ }\textbf {\bibinfo {volume} {77}},\ \bibinfo {pages}
  {4744--4747} (\bibinfo {year} {1996})}\BibitemShut {NoStop}%
\bibitem [{\citenamefont {Batisti{\'{c}}}\ and\ \citenamefont
  {Robnik}(2010)}]{Batistic2010}%
  \BibitemOpen
  \bibfield  {author} {\bibinfo {author} {\bibfnamefont {Benjamin}\
  \bibnamefont {Batisti{\'{c}}}}\ and\ \bibinfo {author} {\bibfnamefont
  {Marko}\ \bibnamefont {Robnik}},\ }\bibfield  {title} {\enquote {\bibinfo
  {title} {Semiempirical theory of level spacing distribution beyond the
  {B}erry{\textendash}{R}obnik regime: modeling the localization and the
  tunneling effects},}\ }\href {\doibase 10.1088/1751-8113/43/21/215101}
  {\bibfield  {journal} {\bibinfo  {journal} {Jour. Phys. A}\ }\textbf
  {\bibinfo {volume} {43}},\ \bibinfo {pages} {215101} (\bibinfo {year}
  {2010})}\BibitemShut {NoStop}%
\bibitem [{\citenamefont {Batisti{\'{c}}}\ and\ \citenamefont
  {Robnik}(2013{\natexlab{a}})}]{Batistic2013a}%
  \BibitemOpen
  \bibfield  {author} {\bibinfo {author} {\bibfnamefont {Benjamin}\
  \bibnamefont {Batisti{\'{c}}}}\ and\ \bibinfo {author} {\bibfnamefont
  {Marko}\ \bibnamefont {Robnik}},\ }\bibfield  {title} {\enquote {\bibinfo
  {title} {Dynamical localization of chaotic eigenstates in the mixed-type
  systems: spectral statistics in a billiard system after separation of regular
  and chaotic eigenstates},}\ }\href {\doibase 10.1088/1751-8113/46/31/315102}
  {\bibfield  {journal} {\bibinfo  {journal} {Journal of Physics A:
  Mathematical and Theoretical}\ }\textbf {\bibinfo {volume} {46}},\ \bibinfo
  {pages} {315102} (\bibinfo {year} {2013}{\natexlab{a}})}\BibitemShut
  {NoStop}%
\bibitem [{\citenamefont {Batisti{\'{c}}}\ and\ \citenamefont
  {Robnik}(2013{\natexlab{b}})}]{Batistic2013b}%
  \BibitemOpen
  \bibfield  {author} {\bibinfo {author} {\bibfnamefont {Benjamin}\
  \bibnamefont {Batisti{\'{c}}}}\ and\ \bibinfo {author} {\bibfnamefont
  {Marko}\ \bibnamefont {Robnik}},\ }\bibfield  {title} {\enquote {\bibinfo
  {title} {Quantum localization of chaotic eigenstates and the level spacing
  distribution},}\ }\href {\doibase 10.1103/PhysRevE.88.052913} {\bibfield
  {journal} {\bibinfo  {journal} {Phys. Rev. E}\ }\textbf {\bibinfo {volume}
  {88}},\ \bibinfo {pages} {052913} (\bibinfo {year}
  {2013}{\natexlab{b}})}\BibitemShut {NoStop}%
\bibitem [{\citenamefont {Batisti{\'{c}}}\ \emph {et~al.}(2019)\citenamefont
  {Batisti{\'{c}}}, \citenamefont {Lozej},\ and\ \citenamefont
  {Robnik}}]{Batistic2019}%
  \BibitemOpen
  \bibfield  {author} {\bibinfo {author} {\bibfnamefont {Benjamin}\
  \bibnamefont {Batisti{\'{c}}}}, \bibinfo {author} {\bibfnamefont {\ifmmode
  \check{C}\else~\v{C}\fi{}rt}\ \bibnamefont {Lozej}}, \ and\ \bibinfo {author}
  {\bibfnamefont {Marko}\ \bibnamefont {Robnik}},\ }\bibfield  {title}
  {\enquote {\bibinfo {title} {Statistical properties of the localization
  measure of chaotic eigenstates and the spectral statistics in a mixed-type
  billiard},}\ }\href {\doibase 10.1103/PhysRevE.100.062208} {\bibfield
  {journal} {\bibinfo  {journal} {Phys. Rev. E}\ }\textbf {\bibinfo {volume}
  {100}},\ \bibinfo {pages} {062208} (\bibinfo {year} {2019})}\BibitemShut
  {NoStop}%
\bibitem [{\citenamefont {Robnik}(2020)}]{Robnik2020}%
  \BibitemOpen
  \bibfield  {author} {\bibinfo {author} {\bibfnamefont {Marko}\ \bibnamefont
  {Robnik}},\ }\enquote {\bibinfo {title} {Recent advances in quantum chaos of
  generic systems},}\ in\ \href {\doibase 10.1007/978-1-0716-0421-2_730} {\emph
  {\bibinfo {booktitle} {Synergetics}}},\ \bibinfo {editor} {edited by\
  \bibinfo {editor} {\bibfnamefont {Axel}\ \bibnamefont {Hutt}}\ and\ \bibinfo
  {editor} {\bibfnamefont {Hermann}\ \bibnamefont {Haken}}}\ (\bibinfo
  {publisher} {Springer US},\ \bibinfo {year} {2020})\ pp.\ \bibinfo {pages}
  {133--148}\BibitemShut {NoStop}%
\bibitem [{\citenamefont {Rozenbaum}\ and\ \citenamefont
  {Galitski}(2017)}]{Rozenbaum2017a}%
  \BibitemOpen
  \bibfield  {author} {\bibinfo {author} {\bibfnamefont {Efim~B.}\ \bibnamefont
  {Rozenbaum}}\ and\ \bibinfo {author} {\bibfnamefont {Victor}\ \bibnamefont
  {Galitski}},\ }\bibfield  {title} {\enquote {\bibinfo {title} {Dynamical
  localization of coupled relativistic kicked rotors},}\ }\href {\doibase
  10.1103/PhysRevB.95.064303} {\bibfield  {journal} {\bibinfo  {journal} {Phys.
  Rev. B}\ }\textbf {\bibinfo {volume} {95}},\ \bibinfo {pages} {064303}
  (\bibinfo {year} {2017})}\BibitemShut {NoStop}%
\bibitem [{\citenamefont {Fava}\ \emph {et~al.}(2020)\citenamefont {Fava},
  \citenamefont {Fazio},\ and\ \citenamefont {Russomanno}}]{Fava2020}%
  \BibitemOpen
  \bibfield  {author} {\bibinfo {author} {\bibfnamefont {Michele}\ \bibnamefont
  {Fava}}, \bibinfo {author} {\bibfnamefont {Rosario}\ \bibnamefont {Fazio}}, \
  and\ \bibinfo {author} {\bibfnamefont {Angelo}\ \bibnamefont {Russomanno}},\
  }\bibfield  {title} {\enquote {\bibinfo {title} {Many-body dynamical
  localization in the kicked {B}ose-{H}ubbard chain},}\ }\href {\doibase
  10.1103/PhysRevB.101.064302} {\bibfield  {journal} {\bibinfo  {journal}
  {Phys. Rev. B}\ }\textbf {\bibinfo {volume} {101}},\ \bibinfo {pages}
  {064302} (\bibinfo {year} {2020})}\BibitemShut {NoStop}%
\bibitem [{\citenamefont {Rylands}\ \emph {et~al.}(2020)\citenamefont
  {Rylands}, \citenamefont {Rozenbaum}, \citenamefont {Galitski},\ and\
  \citenamefont {Konik}}]{Rylands2020}%
  \BibitemOpen
  \bibfield  {author} {\bibinfo {author} {\bibfnamefont {Colin}\ \bibnamefont
  {Rylands}}, \bibinfo {author} {\bibfnamefont {Efim~B.}\ \bibnamefont
  {Rozenbaum}}, \bibinfo {author} {\bibfnamefont {Victor}\ \bibnamefont
  {Galitski}}, \ and\ \bibinfo {author} {\bibfnamefont {Robert}\ \bibnamefont
  {Konik}},\ }\bibfield  {title} {\enquote {\bibinfo {title} {Many-body
  dynamical localization in a kicked {L}ieb-{L}iniger gas},}\ }\href {\doibase
  10.1103/PhysRevLett.124.155302} {\bibfield  {journal} {\bibinfo  {journal}
  {Phys. Rev. Lett.}\ }\textbf {\bibinfo {volume} {124}},\ \bibinfo {pages}
  {155302} (\bibinfo {year} {2020})}\BibitemShut {NoStop}%
\bibitem [{\citenamefont {Campbell}(1966)}]{Campbell1966}%
  \BibitemOpen
  \bibfield  {author} {\bibinfo {author} {\bibfnamefont {L~L}\ \bibnamefont
  {Campbell}},\ }\bibfield  {title} {\enquote {\bibinfo {title} {Exponential
  entropy as a measure of extent of a distribution},}\ }\href {\doibase
  10.1007/BF00533058} {\bibfield  {journal} {\bibinfo  {journal} {Z.
  Wahrscheinlichkeitstheorie verw Gebiete}\ }\textbf {\bibinfo {volume} {5}},\
  \bibinfo {pages} {217--225} (\bibinfo {year} {1966})}\BibitemShut {NoStop}%
\bibitem [{\citenamefont {Jost}(2006)}]{Jost2006}%
  \BibitemOpen
  \bibfield  {author} {\bibinfo {author} {\bibfnamefont {Lou}\ \bibnamefont
  {Jost}},\ }\bibfield  {title} {\enquote {\bibinfo {title} {Entropy and
  diversity},}\ }\href {\doibase 10.1111/j.2006.0030-1299.14714.x} {\bibfield
  {journal} {\bibinfo  {journal} {Oikos}\ }\textbf {\bibinfo {volume} {113}},\
  \bibinfo {pages} {363--375} (\bibinfo {year} {2006})}\BibitemShut {NoStop}%
\bibitem [{\citenamefont {Jelinek}\ \emph {et~al.}(1977)\citenamefont
  {Jelinek}, \citenamefont {Mercer}, \citenamefont {Bahl},\ and\ \citenamefont
  {Baker}}]{Jelinek1977}%
  \BibitemOpen
  \bibfield  {author} {\bibinfo {author} {\bibfnamefont {F.}~\bibnamefont
  {Jelinek}}, \bibinfo {author} {\bibfnamefont {R.~L.}\ \bibnamefont {Mercer}},
  \bibinfo {author} {\bibfnamefont {L.~R.}\ \bibnamefont {Bahl}}, \ and\
  \bibinfo {author} {\bibfnamefont {J.~K.}\ \bibnamefont {Baker}},\ }\bibfield
  {title} {\enquote {\bibinfo {title} {{P}erplexity{\textemdash}a measure of
  the difficulty of speech recognition tasks},}\ }\href {\doibase
  10.1121/1.2016299} {\bibfield  {journal} {\bibinfo  {journal} {Jour. Acou.
  Soc. Am.}\ }\textbf {\bibinfo {volume} {62}},\ \bibinfo {pages} {S63--S63}
  (\bibinfo {year} {1977})}\BibitemShut {NoStop}%
\bibitem [{\citenamefont {Brown}\ \emph {et~al.}(1992)\citenamefont {Brown},
  \citenamefont {Pietra}, \citenamefont {Mercer}, \citenamefont {Pietra},\ and\
  \citenamefont {Lai}}]{Brown1992}%
  \BibitemOpen
  \bibfield  {author} {\bibinfo {author} {\bibfnamefont {Peter~F.}\
  \bibnamefont {Brown}}, \bibinfo {author} {\bibfnamefont {Vincent J.~Della}\
  \bibnamefont {Pietra}}, \bibinfo {author} {\bibfnamefont {Robert~L.}\
  \bibnamefont {Mercer}}, \bibinfo {author} {\bibfnamefont {Stephen A.~Della}\
  \bibnamefont {Pietra}}, \ and\ \bibinfo {author} {\bibfnamefont
  {Jennifer~C.}\ \bibnamefont {Lai}},\ }\bibfield  {title} {\enquote {\bibinfo
  {title} {An estimate of an upper bound for the entropy of {E}nglish},}\
  }\href {https://dl.acm.org/doi/10.5555/146680.146685} {\bibfield  {journal}
  {\bibinfo  {journal} {Comp. Ling.}\ }\textbf {\bibinfo {volume} {18}}
  (\bibinfo {year} {1992})}\BibitemShut {NoStop}%
\bibitem [{\citenamefont {Gorin}\ \emph {et~al.}(1997)\citenamefont {Gorin},
  \citenamefont {Korsch},\ and\ \citenamefont {Mirbach}}]{Gorin1997}%
  \BibitemOpen
  \bibfield  {author} {\bibinfo {author} {\bibfnamefont {T.}~\bibnamefont
  {Gorin}}, \bibinfo {author} {\bibfnamefont {H.J.}\ \bibnamefont {Korsch}}, \
  and\ \bibinfo {author} {\bibfnamefont {B.}~\bibnamefont {Mirbach}},\
  }\bibfield  {title} {\enquote {\bibinfo {title} {Phase-space localization and
  level spacing distributions for a driven rotor with mixed regular/chaotic
  dynamics},}\ }\href {\doibase https://doi.org/10.1016/S0301-0104(97)00036-0}
  {\bibfield  {journal} {\bibinfo  {journal} {Chem. Phys.}\ }\textbf {\bibinfo
  {volume} {217}},\ \bibinfo {pages} {145 -- 153} (\bibinfo {year}
  {1997})}\BibitemShut {NoStop}%
\bibitem [{\citenamefont {Husimi}(1940)}]{Husimi1940}%
  \BibitemOpen
  \bibfield  {author} {\bibinfo {author} {\bibfnamefont {K\^odi}\ \bibnamefont
  {Husimi}},\ }\bibfield  {title} {\enquote {\bibinfo {title} {{Some Formal
  Properties of the Density Matrix}},}\ }\href {\doibase
  10.11429/ppmsj1919.22.4_264} {\bibfield  {journal} {\bibinfo  {journal}
  {{Proceedings of the Physico-Mathematical Society of Japan. 3rd Series}}\
  }\textbf {\bibinfo {volume} {22}},\ \bibinfo {pages} {264--314} (\bibinfo
  {year} {1940})}\BibitemShut {NoStop}%
\bibitem [{\citenamefont {Wehrl}(1978)}]{Wehrl1978}%
  \BibitemOpen
  \bibfield  {author} {\bibinfo {author} {\bibfnamefont {Alfred}\ \bibnamefont
  {Wehrl}},\ }\bibfield  {title} {\enquote {\bibinfo {title} {General
  properties of entropy},}\ }\href {\doibase 10.1103/RevModPhys.50.221}
  {\bibfield  {journal} {\bibinfo  {journal} {Rev. Mod. Phys.}\ }\textbf
  {\bibinfo {volume} {50}},\ \bibinfo {pages} {221--260} (\bibinfo {year}
  {1978})}\BibitemShut {NoStop}%
\bibitem [{\citenamefont {Gnutzmann}\ and\ \citenamefont
  {Zyczkowski}(2001)}]{Gnutzmann2001}%
  \BibitemOpen
  \bibfield  {author} {\bibinfo {author} {\bibfnamefont {Sven}\ \bibnamefont
  {Gnutzmann}}\ and\ \bibinfo {author} {\bibfnamefont {Karol}\ \bibnamefont
  {Zyczkowski}},\ }\bibfield  {title} {\enquote {\bibinfo {title}
  {R{\'{e}}nyi-{W}ehrl entropies as measures of localization in phase space},}\
  }\href {\doibase 10.1088/0305-4470/34/47/317} {\bibfield  {journal} {\bibinfo
   {journal} {J. Phys. A}\ }\textbf {\bibinfo {volume} {34}},\ \bibinfo {pages}
  {10123--10139} (\bibinfo {year} {2001})}\BibitemShut {NoStop}%
\bibitem [{\citenamefont {Wang}\ and\ \citenamefont {Robnik}(2020)}]{Wang2020}%
  \BibitemOpen
  \bibfield  {author} {\bibinfo {author} {\bibfnamefont {Qian}\ \bibnamefont
  {Wang}}\ and\ \bibinfo {author} {\bibfnamefont {Marko}\ \bibnamefont
  {Robnik}},\ }\bibfield  {title} {\enquote {\bibinfo {title} {Statistical
  properties of the localization measure of chaotic eigenstates in the {D}icke
  model},}\ }\href {\doibase 10.1103/PhysRevE.102.032212} {\bibfield  {journal}
  {\bibinfo  {journal} {Phys. Rev. E}\ }\textbf {\bibinfo {volume} {102}},\
  \bibinfo {pages} {032212} (\bibinfo {year} {2020})}\BibitemShut {NoStop}%
\bibitem [{\citenamefont {Pilatowsky-Cameo}\ \emph
  {et~al.}(2021{\natexlab{a}})\citenamefont {Pilatowsky-Cameo}, \citenamefont
  {Villase{\~{n}}or}, \citenamefont {Bastarrachea-Magnani}, \citenamefont
  {Lerma-Hern{\'{a}}ndez}, \citenamefont {Santos},\ and\ \citenamefont
  {Hirsch}}]{Pilatowsky2021NatCommun}%
  \BibitemOpen
  \bibfield  {author} {\bibinfo {author} {\bibfnamefont {Sa{\'{u}}l}\
  \bibnamefont {Pilatowsky-Cameo}}, \bibinfo {author} {\bibfnamefont {David}\
  \bibnamefont {Villase{\~{n}}or}}, \bibinfo {author} {\bibfnamefont
  {Miguel~A.}\ \bibnamefont {Bastarrachea-Magnani}}, \bibinfo {author}
  {\bibfnamefont {Sergio}\ \bibnamefont {Lerma-Hern{\'{a}}ndez}}, \bibinfo
  {author} {\bibfnamefont {Lea~F.}\ \bibnamefont {Santos}}, \ and\ \bibinfo
  {author} {\bibfnamefont {Jorge~G.}\ \bibnamefont {Hirsch}},\ }\bibfield
  {title} {\enquote {\bibinfo {title} {Ubiquitous quantum scarring does not
  prevent ergodicity},}\ }\href {\doibase 10.1038/s41467-021-21123-5}
  {\bibfield  {journal} {\bibinfo  {journal} {Nat. Commun.}\ }\textbf {\bibinfo
  {volume} {12}} (\bibinfo {year} {2021}{\natexlab{a}}),\
  10.1038/s41467-021-21123-5}\BibitemShut {NoStop}%
\bibitem [{\citenamefont {Dicke}(1954)}]{Dicke1954}%
  \BibitemOpen
  \bibfield  {author} {\bibinfo {author} {\bibfnamefont {R.~H.}\ \bibnamefont
  {Dicke}},\ }\bibfield  {title} {\enquote {\bibinfo {title} {Coherence in
  spontaneous radiation processes},}\ }\href {\doibase 10.1103/PhysRev.93.99}
  {\bibfield  {journal} {\bibinfo  {journal} {Phys. Rev.}\ }\textbf {\bibinfo
  {volume} {93}},\ \bibinfo {pages} {99} (\bibinfo {year} {1954})}\BibitemShut
  {NoStop}%
\bibitem [{\citenamefont {Hepp}\ and\ \citenamefont
  {Lieb}(1973{\natexlab{a}})}]{Hepp1973a}%
  \BibitemOpen
  \bibfield  {author} {\bibinfo {author} {\bibfnamefont {Klaus}\ \bibnamefont
  {Hepp}}\ and\ \bibinfo {author} {\bibfnamefont {Elliott~H}\ \bibnamefont
  {Lieb}},\ }\bibfield  {title} {\enquote {\bibinfo {title} {On the
  superradiant phase transition for molecules in a quantized radiation field:
  the {D}icke maser model},}\ }\href {\doibase
  https://doi.org/10.1016/0003-4916(73)90039-0} {\bibfield  {journal} {\bibinfo
   {journal} {Ann. Phys. (N.Y.)}\ }\textbf {\bibinfo {volume} {76}},\ \bibinfo
  {pages} {360 -- 404} (\bibinfo {year} {1973}{\natexlab{a}})}\BibitemShut
  {NoStop}%
\bibitem [{\citenamefont {Wang}\ and\ \citenamefont {Hioe}(1973)}]{Wang1973}%
  \BibitemOpen
  \bibfield  {author} {\bibinfo {author} {\bibfnamefont {Y.~K.}\ \bibnamefont
  {Wang}}\ and\ \bibinfo {author} {\bibfnamefont {F.~T.}\ \bibnamefont
  {Hioe}},\ }\bibfield  {title} {\enquote {\bibinfo {title} {Phase transition
  in the {D}icke model of superradiance},}\ }\href {\doibase
  10.1103/PhysRevA.7.831} {\bibfield  {journal} {\bibinfo  {journal} {Phys.
  Rev. A}\ }\textbf {\bibinfo {volume} {7}},\ \bibinfo {pages} {831--836}
  (\bibinfo {year} {1973})}\BibitemShut {NoStop}%
\bibitem [{\citenamefont {Garraway}(2011)}]{Garraway2011}%
  \BibitemOpen
  \bibfield  {author} {\bibinfo {author} {\bibfnamefont {Barry~M.}\
  \bibnamefont {Garraway}},\ }\bibfield  {title} {\enquote {\bibinfo {title}
  {The {D}icke model in quantum optics: {D}icke model revisited},}\ }\href
  {\doibase 10.1098/rsta.2010.0333} {\bibfield  {journal} {\bibinfo  {journal}
  {Philos. Trans. Royal Soc. A}\ }\textbf {\bibinfo {volume} {369}},\ \bibinfo
  {pages} {1137} (\bibinfo {year} {2011})}\BibitemShut {NoStop}%
\bibitem [{\citenamefont {Kirton}\ \emph {et~al.}(2019)\citenamefont {Kirton},
  \citenamefont {Roses}, \citenamefont {Keeling},\ and\ \citenamefont
  {Dalla~Torre}}]{Kirton2019}%
  \BibitemOpen
  \bibfield  {author} {\bibinfo {author} {\bibfnamefont {Peter}\ \bibnamefont
  {Kirton}}, \bibinfo {author} {\bibfnamefont {Mor~M.}\ \bibnamefont {Roses}},
  \bibinfo {author} {\bibfnamefont {Jonathan}\ \bibnamefont {Keeling}}, \ and\
  \bibinfo {author} {\bibfnamefont {Emanuele~G.}\ \bibnamefont {Dalla~Torre}},\
  }\bibfield  {title} {\enquote {\bibinfo {title} {Introduction to the {D}icke
  model: From equilibrium to nonequilibrium, and vice versa},}\ }\href
  {\doibase 10.1002/qute.201800043} {\bibfield  {journal} {\bibinfo  {journal}
  {Adv. Quantum Technol.}\ }\textbf {\bibinfo {volume} {2}},\ \bibinfo {pages}
  {1800043} (\bibinfo {year} {2019})}\BibitemShut {NoStop}%
\bibitem [{\citenamefont {Ch\'avez-Carlos}\ \emph {et~al.}(2019)\citenamefont
  {Ch\'avez-Carlos}, \citenamefont {L\'opez-del Carpio}, \citenamefont
  {Bastarrachea-Magnani}, \citenamefont {Str\'ansk\'y}, \citenamefont
  {Lerma-Hern\'andez}, \citenamefont {Santos},\ and\ \citenamefont
  {Hirsch}}]{Chavez2019}%
  \BibitemOpen
  \bibfield  {author} {\bibinfo {author} {\bibfnamefont {Jorge}\ \bibnamefont
  {Ch\'avez-Carlos}}, \bibinfo {author} {\bibfnamefont {B.}~\bibnamefont
  {L\'opez-del Carpio}}, \bibinfo {author} {\bibfnamefont {Miguel~A.}\
  \bibnamefont {Bastarrachea-Magnani}}, \bibinfo {author} {\bibfnamefont
  {Pavel}\ \bibnamefont {Str\'ansk\'y}}, \bibinfo {author} {\bibfnamefont
  {Sergio}\ \bibnamefont {Lerma-Hern\'andez}}, \bibinfo {author} {\bibfnamefont
  {Lea~F.}\ \bibnamefont {Santos}}, \ and\ \bibinfo {author} {\bibfnamefont
  {Jorge~G.}\ \bibnamefont {Hirsch}},\ }\bibfield  {title} {\enquote {\bibinfo
  {title} {Quantum and classical {L}yapunov exponents in atom-field interaction
  systems},}\ }\href {\doibase 10.1103/PhysRevLett.122.024101} {\bibfield
  {journal} {\bibinfo  {journal} {Phys. Rev. Lett.}\ }\textbf {\bibinfo
  {volume} {122}},\ \bibinfo {pages} {024101} (\bibinfo {year}
  {2019})}\BibitemShut {NoStop}%
\bibitem [{\citenamefont {Lewis-Swan}\ \emph {et~al.}(2019)\citenamefont
  {Lewis-Swan}, \citenamefont {Safavi-Naini}, \citenamefont {Bollinger},\ and\
  \citenamefont {Rey}}]{Lewis-Swan2019}%
  \BibitemOpen
  \bibfield  {author} {\bibinfo {author} {\bibfnamefont {R.~J.}\ \bibnamefont
  {Lewis-Swan}}, \bibinfo {author} {\bibfnamefont {A.}~\bibnamefont
  {Safavi-Naini}}, \bibinfo {author} {\bibfnamefont {J.~J.}\ \bibnamefont
  {Bollinger}}, \ and\ \bibinfo {author} {\bibfnamefont {A.~M.}\ \bibnamefont
  {Rey}},\ }\bibfield  {title} {\enquote {\bibinfo {title} {Unifying ,
  thermalization and entanglement through measurement of fidelity
  out-of-time-order correlators in the {D}icke model},}\ }\href {\doibase
  10.1038/s41467-019-09436-y} {\bibfield  {journal} {\bibinfo  {journal} {Nat.
  Comm.}\ }\textbf {\bibinfo {volume} {10}},\ \bibinfo {pages} {1581} (\bibinfo
  {year} {2019})}\BibitemShut {NoStop}%
\bibitem [{\citenamefont {Pilatowsky-Cameo}\ \emph {et~al.}(2020)\citenamefont
  {Pilatowsky-Cameo}, \citenamefont {Ch\'avez-Carlos}, \citenamefont
  {Bastarrachea-Magnani}, \citenamefont {Str\'ansk\'y}, \citenamefont
  {Lerma-Hern\'andez}, \citenamefont {Santos},\ and\ \citenamefont
  {Hirsch}}]{Pilatowsky2020}%
  \BibitemOpen
  \bibfield  {author} {\bibinfo {author} {\bibfnamefont {Sa\'ul}\ \bibnamefont
  {Pilatowsky-Cameo}}, \bibinfo {author} {\bibfnamefont {Jorge}\ \bibnamefont
  {Ch\'avez-Carlos}}, \bibinfo {author} {\bibfnamefont {Miguel~A.}\
  \bibnamefont {Bastarrachea-Magnani}}, \bibinfo {author} {\bibfnamefont
  {Pavel}\ \bibnamefont {Str\'ansk\'y}}, \bibinfo {author} {\bibfnamefont
  {Sergio}\ \bibnamefont {Lerma-Hern\'andez}}, \bibinfo {author} {\bibfnamefont
  {Lea~F.}\ \bibnamefont {Santos}}, \ and\ \bibinfo {author} {\bibfnamefont
  {Jorge~G.}\ \bibnamefont {Hirsch}},\ }\bibfield  {title} {\enquote {\bibinfo
  {title} {Positive quantum {L}yapunov exponents in experimental systems with a
  regular classical limit},}\ }\href {\doibase 10.1103/PhysRevE.101.010202}
  {\bibfield  {journal} {\bibinfo  {journal} {Phys. Rev. E}\ }\textbf {\bibinfo
  {volume} {101}},\ \bibinfo {pages} {010202(R)} (\bibinfo {year}
  {2020})}\BibitemShut {NoStop}%
\bibitem [{\citenamefont {de~Aguiar}\ \emph {et~al.}(1992)\citenamefont
  {de~Aguiar}, \citenamefont {Furuya}, \citenamefont {Lewenkopf},\ and\
  \citenamefont {Nemes}}]{Deaguiar1992}%
  \BibitemOpen
  \bibfield  {author} {\bibinfo {author} {\bibfnamefont {M.A.M}\ \bibnamefont
  {de~Aguiar}}, \bibinfo {author} {\bibfnamefont {K}~\bibnamefont {Furuya}},
  \bibinfo {author} {\bibfnamefont {C.H}\ \bibnamefont {Lewenkopf}}, \ and\
  \bibinfo {author} {\bibfnamefont {M.C}\ \bibnamefont {Nemes}},\ }\bibfield
  {title} {\enquote {\bibinfo {title} {Chaos in a spin-boson system: Classical
  analysis},}\ }\href {\doibase https://doi.org/10.1016/0003-4916(92)90178-O}
  {\bibfield  {journal} {\bibinfo  {journal} {Ann. Phys.}\ }\textbf {\bibinfo
  {volume} {216}},\ \bibinfo {pages} {291 -- 312} (\bibinfo {year}
  {1992})}\BibitemShut {NoStop}%
\bibitem [{\citenamefont {Furuya}\ \emph {et~al.}(1992)\citenamefont {Furuya},
  \citenamefont {de~Aguiar}, \citenamefont {Lewenkopf},\ and\ \citenamefont
  {Nemes}}]{Furuya1992}%
  \BibitemOpen
  \bibfield  {author} {\bibinfo {author} {\bibfnamefont {K}~\bibnamefont
  {Furuya}}, \bibinfo {author} {\bibfnamefont {M.A.M}\ \bibnamefont
  {de~Aguiar}}, \bibinfo {author} {\bibfnamefont {C.H}\ \bibnamefont
  {Lewenkopf}}, \ and\ \bibinfo {author} {\bibfnamefont {M.C}\ \bibnamefont
  {Nemes}},\ }\bibfield  {title} {\enquote {\bibinfo {title} {{H}usimi
  distributions of a spin-boson system and the signatures of its classical
  dynamics},}\ }\href {\doibase 10.1016/0003-4916(92)90179-p} {\bibfield
  {journal} {\bibinfo  {journal} {Ann. of Phys.}\ }\textbf {\bibinfo {volume}
  {216}},\ \bibinfo {pages} {313--322} (\bibinfo {year} {1992})}\BibitemShut
  {NoStop}%
\bibitem [{\citenamefont {Bakemeier}\ \emph {et~al.}(2013)\citenamefont
  {Bakemeier}, \citenamefont {Alvermann},\ and\ \citenamefont
  {Fehske}}]{Bakemeier2013}%
  \BibitemOpen
  \bibfield  {author} {\bibinfo {author} {\bibfnamefont {L.}~\bibnamefont
  {Bakemeier}}, \bibinfo {author} {\bibfnamefont {A.}~\bibnamefont
  {Alvermann}}, \ and\ \bibinfo {author} {\bibfnamefont {H.}~\bibnamefont
  {Fehske}},\ }\bibfield  {title} {\enquote {\bibinfo {title} {Dynamics of the
  {D}icke model close to the classical limit},}\ }\href {\doibase
  10.1103/PhysRevA.88.043835} {\bibfield  {journal} {\bibinfo  {journal} {Phys.
  Rev. A}\ }\textbf {\bibinfo {volume} {88}},\ \bibinfo {pages} {043835}
  (\bibinfo {year} {2013})}\BibitemShut {NoStop}%
\bibitem [{\citenamefont {Pilatowsky-Cameo}\ \emph
  {et~al.}(2021{\natexlab{b}})\citenamefont {Pilatowsky-Cameo}, \citenamefont
  {Villase{\~{n}}or}, \citenamefont {Bastarrachea-Magnani}, \citenamefont
  {Lerma-Hern\'andez}, \citenamefont {Santos},\ and\ \citenamefont
  {Hirsch}}]{Pilatowsky2021}%
  \BibitemOpen
  \bibfield  {author} {\bibinfo {author} {\bibfnamefont {Sa\'ul}\ \bibnamefont
  {Pilatowsky-Cameo}}, \bibinfo {author} {\bibfnamefont {David}\ \bibnamefont
  {Villase{\~{n}}or}}, \bibinfo {author} {\bibfnamefont {Miguel~A.}\
  \bibnamefont {Bastarrachea-Magnani}}, \bibinfo {author} {\bibfnamefont
  {Sergio}\ \bibnamefont {Lerma-Hern\'andez}}, \bibinfo {author} {\bibfnamefont
  {Lea~F.}\ \bibnamefont {Santos}}, \ and\ \bibinfo {author} {\bibfnamefont
  {Jorge~G.}\ \bibnamefont {Hirsch}},\ }\bibfield  {title} {\enquote {\bibinfo
  {title} {Quantum scarring in a spin-boson system: fundamental families of
  periodic orbits},}\ }\href {\doibase 10.1088/1367-2630/abd2e6} {\bibfield
  {journal} {\bibinfo  {journal} {New J. of Phys.}\ } (\bibinfo {year}
  {2021}{\natexlab{b}}),\ 10.1088/1367-2630/abd2e6},\ \bibinfo {note} {({I}n
  press)}\BibitemShut {NoStop}%
\bibitem [{\citenamefont {Altland}\ and\ \citenamefont
  {Haake}(2012)}]{Altland2012NJP}%
  \BibitemOpen
  \bibfield  {author} {\bibinfo {author} {\bibfnamefont {Alexander}\
  \bibnamefont {Altland}}\ and\ \bibinfo {author} {\bibfnamefont {Fritz}\
  \bibnamefont {Haake}},\ }\bibfield  {title} {\enquote {\bibinfo {title}
  {Equilibration and macroscopic quantum fluctuations in the {D}icke model},}\
  }\href {http://stacks.iop.org/1367-2630/14/i=7/a=073011} {\bibfield
  {journal} {\bibinfo  {journal} {New J. Phys.}\ }\textbf {\bibinfo {volume}
  {14}},\ \bibinfo {pages} {073011} (\bibinfo {year} {2012})}\BibitemShut
  {NoStop}%
\bibitem [{\citenamefont {Kloc}\ \emph {et~al.}(2018)\citenamefont {Kloc},
  \citenamefont {Str\'ansk\'y},\ and\ \citenamefont {Cejnar}}]{Kloc2018}%
  \BibitemOpen
  \bibfield  {author} {\bibinfo {author} {\bibfnamefont {Michal}\ \bibnamefont
  {Kloc}}, \bibinfo {author} {\bibfnamefont {Pavel}\ \bibnamefont
  {Str\'ansk\'y}}, \ and\ \bibinfo {author} {\bibfnamefont {Pavel}\
  \bibnamefont {Cejnar}},\ }\bibfield  {title} {\enquote {\bibinfo {title}
  {Quantum quench dynamics in {D}icke superradiance models},}\ }\href {\doibase
  10.1103/PhysRevA.98.013836} {\bibfield  {journal} {\bibinfo  {journal} {Phys.
  Rev. A}\ }\textbf {\bibinfo {volume} {98}},\ \bibinfo {pages} {013836}
  (\bibinfo {year} {2018})}\BibitemShut {NoStop}%
\bibitem [{\citenamefont {Lerma-Hern\'andez}\ \emph {et~al.}(2018)\citenamefont
  {Lerma-Hern\'andez}, \citenamefont {Ch\'avez-Carlos}, \citenamefont
  {Bastarrachea-Magnani}, \citenamefont {Santos},\ and\ \citenamefont
  {Hirsch}}]{Lerma2018}%
  \BibitemOpen
  \bibfield  {author} {\bibinfo {author} {\bibfnamefont {Sergio}\ \bibnamefont
  {Lerma-Hern\'andez}}, \bibinfo {author} {\bibfnamefont {Jorge}\ \bibnamefont
  {Ch\'avez-Carlos}}, \bibinfo {author} {\bibfnamefont {Miguel~A.}\
  \bibnamefont {Bastarrachea-Magnani}}, \bibinfo {author} {\bibfnamefont
  {Lea~F.}\ \bibnamefont {Santos}}, \ and\ \bibinfo {author} {\bibfnamefont
  {Jorge~G.}\ \bibnamefont {Hirsch}},\ }\bibfield  {title} {\enquote {\bibinfo
  {title} {Analytical description of the survival probability of coherent
  states in regular regimes},}\ }\href
  {http://stacks.iop.org/1751-8121/51/i=47/a=475302} {\bibfield  {journal}
  {\bibinfo  {journal} {J. Phys. A}\ }\textbf {\bibinfo {volume} {51}},\
  \bibinfo {pages} {475302} (\bibinfo {year} {2018})}\BibitemShut {NoStop}%
\bibitem [{\citenamefont {Lerma-Hern\'andez}\ \emph {et~al.}(2019)\citenamefont
  {Lerma-Hern\'andez}, \citenamefont {Villase\~nor}, \citenamefont
  {Bastarrachea-Magnani}, \citenamefont {Torres-Herrera}, \citenamefont
  {Santos},\ and\ \citenamefont {Hirsch}}]{Lerma2019}%
  \BibitemOpen
  \bibfield  {author} {\bibinfo {author} {\bibfnamefont {S.}~\bibnamefont
  {Lerma-Hern\'andez}}, \bibinfo {author} {\bibfnamefont {D.}~\bibnamefont
  {Villase\~nor}}, \bibinfo {author} {\bibfnamefont {M.~A.}\ \bibnamefont
  {Bastarrachea-Magnani}}, \bibinfo {author} {\bibfnamefont {E.~J.}\
  \bibnamefont {Torres-Herrera}}, \bibinfo {author} {\bibfnamefont {L.~F.}\
  \bibnamefont {Santos}}, \ and\ \bibinfo {author} {\bibfnamefont {J.~G.}\
  \bibnamefont {Hirsch}},\ }\bibfield  {title} {\enquote {\bibinfo {title}
  {Dynamical signatures of quantum chaos and relaxation time scales in a
  spin-boson system},}\ }\href {\doibase 10.1103/PhysRevE.100.012218}
  {\bibfield  {journal} {\bibinfo  {journal} {Phys. Rev. E}\ }\textbf {\bibinfo
  {volume} {100}},\ \bibinfo {pages} {012218} (\bibinfo {year}
  {2019})}\BibitemShut {NoStop}%
\bibitem [{\citenamefont {{Villase\~nor}}\ \emph {et~al.}(2020)\citenamefont
  {{Villase\~nor}}, \citenamefont {Pilatowsky-Cameo}, \citenamefont
  {Bastarrachea-Magnani}, \citenamefont {Lerma}, \citenamefont {Santos},\ and\
  \citenamefont {Hirsch}}]{Villasenor2020}%
  \BibitemOpen
  \bibfield  {author} {\bibinfo {author} {\bibfnamefont {David}\ \bibnamefont
  {{Villase\~nor}}}, \bibinfo {author} {\bibfnamefont {Sa\'ul}\ \bibnamefont
  {Pilatowsky-Cameo}}, \bibinfo {author} {\bibfnamefont {Miguel~A}\
  \bibnamefont {Bastarrachea-Magnani}}, \bibinfo {author} {\bibfnamefont
  {Sergio}\ \bibnamefont {Lerma}}, \bibinfo {author} {\bibfnamefont {Lea~F}\
  \bibnamefont {Santos}}, \ and\ \bibinfo {author} {\bibfnamefont {Jorge~G}\
  \bibnamefont {Hirsch}},\ }\bibfield  {title} {\enquote {\bibinfo {title}
  {Quantum vs classical dynamics in a spin-boson system: manifestations of
  spectral correlations and scarring},}\ }\href
  {http://iopscience.iop.org/article/10.1088/1367-2630/ab8ef8} {\bibfield
  {journal} {\bibinfo  {journal} {New J. Phys.}\ }\textbf {\bibinfo {volume}
  {22}},\ \bibinfo {pages} {063036} (\bibinfo {year} {2020})}\BibitemShut
  {NoStop}%
\bibitem [{\citenamefont {Jaako}\ \emph {et~al.}(2016)\citenamefont {Jaako},
  \citenamefont {Xiang}, \citenamefont {Garcia-Ripoll},\ and\ \citenamefont
  {Rabl}}]{Jaako2016}%
  \BibitemOpen
  \bibfield  {author} {\bibinfo {author} {\bibfnamefont {Tuomas}\ \bibnamefont
  {Jaako}}, \bibinfo {author} {\bibfnamefont {Ze-Liang}\ \bibnamefont {Xiang}},
  \bibinfo {author} {\bibfnamefont {Juan~Jos\'e}\ \bibnamefont
  {Garcia-Ripoll}}, \ and\ \bibinfo {author} {\bibfnamefont {Peter}\
  \bibnamefont {Rabl}},\ }\bibfield  {title} {\enquote {\bibinfo {title}
  {Ultrastrong-coupling phenomena beyond the {D}icke model},}\ }\href {\doibase
  10.1103/PhysRevA.94.033850} {\bibfield  {journal} {\bibinfo  {journal} {Phys.
  Rev. A}\ }\textbf {\bibinfo {volume} {94}},\ \bibinfo {pages} {033850}
  (\bibinfo {year} {2016})}\BibitemShut {NoStop}%
\bibitem [{\citenamefont {Baden}\ \emph {et~al.}(2014)\citenamefont {Baden},
  \citenamefont {Arnold}, \citenamefont {Grimsmo}, \citenamefont {Parkins},\
  and\ \citenamefont {Barrett}}]{Baden2014}%
  \BibitemOpen
  \bibfield  {author} {\bibinfo {author} {\bibfnamefont {Markus~P.}\
  \bibnamefont {Baden}}, \bibinfo {author} {\bibfnamefont {Kyle~J.}\
  \bibnamefont {Arnold}}, \bibinfo {author} {\bibfnamefont {Arne~L.}\
  \bibnamefont {Grimsmo}}, \bibinfo {author} {\bibfnamefont {Scott}\
  \bibnamefont {Parkins}}, \ and\ \bibinfo {author} {\bibfnamefont {Murray~D.}\
  \bibnamefont {Barrett}},\ }\bibfield  {title} {\enquote {\bibinfo {title}
  {Realization of the {D}icke model using cavity-assisted {R}aman
  transitions},}\ }\href {\doibase 10.1103/PhysRevLett.113.020408} {\bibfield
  {journal} {\bibinfo  {journal} {Phys. Rev. Lett.}\ }\textbf {\bibinfo
  {volume} {113}},\ \bibinfo {pages} {020408} (\bibinfo {year}
  {2014})}\BibitemShut {NoStop}%
\bibitem [{\citenamefont {Zhang}\ \emph {et~al.}(2018)\citenamefont {Zhang},
  \citenamefont {Lee}, \citenamefont {Kumar}, \citenamefont {Arnold},
  \citenamefont {Masson}, \citenamefont {Grimsmo}, \citenamefont {Parkins},\
  and\ \citenamefont {Barrett}}]{Zhang2018}%
  \BibitemOpen
  \bibfield  {author} {\bibinfo {author} {\bibfnamefont {Zhiqiang}\
  \bibnamefont {Zhang}}, \bibinfo {author} {\bibfnamefont {Chern~Hui}\
  \bibnamefont {Lee}}, \bibinfo {author} {\bibfnamefont {Ravi}\ \bibnamefont
  {Kumar}}, \bibinfo {author} {\bibfnamefont {K.~J.}\ \bibnamefont {Arnold}},
  \bibinfo {author} {\bibfnamefont {Stuart~J.}\ \bibnamefont {Masson}},
  \bibinfo {author} {\bibfnamefont {A.~L.}\ \bibnamefont {Grimsmo}}, \bibinfo
  {author} {\bibfnamefont {A.~S.}\ \bibnamefont {Parkins}}, \ and\ \bibinfo
  {author} {\bibfnamefont {M.~D.}\ \bibnamefont {Barrett}},\ }\bibfield
  {title} {\enquote {\bibinfo {title} {Dicke-model simulation via
  cavity-assisted {R}aman transitions},}\ }\href {\doibase
  10.1103/PhysRevA.97.043858} {\bibfield  {journal} {\bibinfo  {journal} {Phys.
  Rev. A}\ }\textbf {\bibinfo {volume} {97}},\ \bibinfo {pages} {043858}
  (\bibinfo {year} {2018})}\BibitemShut {NoStop}%
\bibitem [{\citenamefont {Cohn}\ \emph {et~al.}(2018)\citenamefont {Cohn},
  \citenamefont {Safavi-Naini}, \citenamefont {Lewis-Swan}, \citenamefont
  {Bohnet}, \citenamefont {G\"arttner}, \citenamefont {Gilmore}, \citenamefont
  {Jordan}, \citenamefont {Rey}, \citenamefont {Bollinger},\ and\ \citenamefont
  {Freericks}}]{Cohn2018}%
  \BibitemOpen
  \bibfield  {author} {\bibinfo {author} {\bibfnamefont {J}~\bibnamefont
  {Cohn}}, \bibinfo {author} {\bibfnamefont {A}~\bibnamefont {Safavi-Naini}},
  \bibinfo {author} {\bibfnamefont {R~J}\ \bibnamefont {Lewis-Swan}}, \bibinfo
  {author} {\bibfnamefont {J~G}\ \bibnamefont {Bohnet}}, \bibinfo {author}
  {\bibfnamefont {M}~\bibnamefont {G\"arttner}}, \bibinfo {author}
  {\bibfnamefont {K~A}\ \bibnamefont {Gilmore}}, \bibinfo {author}
  {\bibfnamefont {J~E}\ \bibnamefont {Jordan}}, \bibinfo {author}
  {\bibfnamefont {A~M}\ \bibnamefont {Rey}}, \bibinfo {author} {\bibfnamefont
  {J~J}\ \bibnamefont {Bollinger}}, \ and\ \bibinfo {author} {\bibfnamefont
  {J~K}\ \bibnamefont {Freericks}},\ }\bibfield  {title} {\enquote {\bibinfo
  {title} {Bang-bang shortcut to adiabaticity in the {D}icke model as realized
  in a penning trap experiment},}\ }\href
  {http://stacks.iop.org/1367-2630/20/i=5/a=055013} {\bibfield  {journal}
  {\bibinfo  {journal} {New J. Phys.}\ }\textbf {\bibinfo {volume} {20}},\
  \bibinfo {pages} {055013} (\bibinfo {year} {2018})}\BibitemShut {NoStop}%
\bibitem [{\citenamefont {Safavi-Naini}\ \emph {et~al.}(2018)\citenamefont
  {Safavi-Naini}, \citenamefont {Lewis-Swan}, \citenamefont {Bohnet},
  \citenamefont {G\"arttner}, \citenamefont {Gilmore}, \citenamefont {Jordan},
  \citenamefont {Cohn}, \citenamefont {Freericks}, \citenamefont {Rey},\ and\
  \citenamefont {Bollinger}}]{Safavi2018}%
  \BibitemOpen
  \bibfield  {author} {\bibinfo {author} {\bibfnamefont {A.}~\bibnamefont
  {Safavi-Naini}}, \bibinfo {author} {\bibfnamefont {R.~J.}\ \bibnamefont
  {Lewis-Swan}}, \bibinfo {author} {\bibfnamefont {J.~G.}\ \bibnamefont
  {Bohnet}}, \bibinfo {author} {\bibfnamefont {M.}~\bibnamefont {G\"arttner}},
  \bibinfo {author} {\bibfnamefont {K.~A.}\ \bibnamefont {Gilmore}}, \bibinfo
  {author} {\bibfnamefont {J.~E.}\ \bibnamefont {Jordan}}, \bibinfo {author}
  {\bibfnamefont {J.}~\bibnamefont {Cohn}}, \bibinfo {author} {\bibfnamefont
  {J.~K.}\ \bibnamefont {Freericks}}, \bibinfo {author} {\bibfnamefont {A.~M.}\
  \bibnamefont {Rey}}, \ and\ \bibinfo {author} {\bibfnamefont {J.~J.}\
  \bibnamefont {Bollinger}},\ }\bibfield  {title} {\enquote {\bibinfo {title}
  {Verification of a many-ion simulator of the {D}icke model through slow
  quenches across a phase transition},}\ }\href {\doibase
  10.1103/PhysRevLett.121.040503} {\bibfield  {journal} {\bibinfo  {journal}
  {Phys. Rev. Lett.}\ }\textbf {\bibinfo {volume} {121}},\ \bibinfo {pages}
  {040503} (\bibinfo {year} {2018})}\BibitemShut {NoStop}%
\bibitem [{\citenamefont {Hall}(1999)}]{Hall1999}%
  \BibitemOpen
  \bibfield  {author} {\bibinfo {author} {\bibfnamefont {Michael J.~W.}\
  \bibnamefont {Hall}},\ }\bibfield  {title} {\enquote {\bibinfo {title}
  {Universal geometric approach to uncertainty, entropy, and information},}\
  }\href {\doibase 10.1103/physreva.59.2602} {\bibfield  {journal} {\bibinfo
  {journal} {Phys. Rev. A}\ }\textbf {\bibinfo {volume} {59}},\ \bibinfo
  {pages} {2602--2615} (\bibinfo {year} {1999})}\BibitemShut {NoStop}%
\bibitem [{\citenamefont {Ott}(2002)}]{OttBook}%
  \BibitemOpen
  \bibfield  {author} {\bibinfo {author} {\bibfnamefont {Edward}\ \bibnamefont
  {Ott}},\ }\href@noop {} {\emph {\bibinfo {title} {Chaos in Dynamical
  Systems}}}\ (\bibinfo  {publisher} {Cambridge University Press},\ \bibinfo
  {address} {Cambridge},\ \bibinfo {year} {2002})\BibitemShut {NoStop}%
\bibitem [{\citenamefont {Murphy}\ \emph {et~al.}(2011)\citenamefont {Murphy},
  \citenamefont {Wortis},\ and\ \citenamefont {Atkinson}}]{Murphy2011}%
  \BibitemOpen
  \bibfield  {author} {\bibinfo {author} {\bibfnamefont {N.~C.}\ \bibnamefont
  {Murphy}}, \bibinfo {author} {\bibfnamefont {R.}~\bibnamefont {Wortis}}, \
  and\ \bibinfo {author} {\bibfnamefont {W.~A.}\ \bibnamefont {Atkinson}},\
  }\bibfield  {title} {\enquote {\bibinfo {title} {Generalized inverse
  participation ratio as a possible measure of localization for interacting
  systems},}\ }\href {\doibase 10.1103/PhysRevB.83.184206} {\bibfield
  {journal} {\bibinfo  {journal} {Phys. Rev. B}\ }\textbf {\bibinfo {volume}
  {83}},\ \bibinfo {pages} {184206} (\bibinfo {year} {2011})}\BibitemShut
  {NoStop}%
\bibitem [{Note1()}]{Note1}%
  \BibitemOpen
  \bibinfo {note} {If the space is discrete, then this probability density
  function is just the probability at each point $\protect \bm {x}_i \in X$,
  $p_i=\varphi (\protect \bm {x}_i)$.}\BibitemShut {Stop}%
\bibitem [{\citenamefont {R\'enyi}(1961)}]{Renyi1961}%
  \BibitemOpen
  \bibfield  {author} {\bibinfo {author} {\bibfnamefont {Alfr\'ed}\
  \bibnamefont {R\'enyi}},\ }\bibfield  {title} {\enquote {\bibinfo {title} {On
  measures of entropy and information},}\ }in\ \href
  {https://projecteuclid.org/euclid.bsmsp/1200512181} {\emph {\bibinfo
  {booktitle} {Proceedings of the Fourth Berkeley Symposium on Mathematical
  Statistics and Probability, Volume 1: Contributions to the Theory of
  Statistics}}}\ (\bibinfo  {publisher} {University of California Press},\
  \bibinfo {year} {1961})\ pp.\ \bibinfo {pages} {547--561}\BibitemShut
  {NoStop}%
\bibitem [{\citenamefont {Shannon}(1948)}]{Shannon1948}%
  \BibitemOpen
  \bibfield  {author} {\bibinfo {author} {\bibfnamefont {Claude~E.}\
  \bibnamefont {Shannon}},\ }\bibfield  {title} {\enquote {\bibinfo {title} {A
  mathematical theory of communication},}\ }\href {\doibase
  10.1002/j.1538-7305.1948.tb01338.x} {\bibfield  {journal} {\bibinfo
  {journal} {Bell Syst. Tech. J.}\ }\textbf {\bibinfo {volume} {27}},\ \bibinfo
  {pages} {379--423} (\bibinfo {year} {1948})}\BibitemShut {NoStop}%
\bibitem [{\citenamefont {Nath}(2020)}]{Nath2020Arxiv}%
  \BibitemOpen
  \bibfield  {author} {\bibinfo {author} {\bibfnamefont {Debraj}\ \bibnamefont
  {Nath}},\ }\href@noop {} {\enquote {\bibinfo {title} {Properties of {R}\'enyi
  complexity ratio of quantum-states: An extension of generalized {R}\'enyi
  complexity},}\ } (\bibinfo {year} {2020}),\ \Eprint
  {http://arxiv.org/abs/2008.05418} {2008.05418} \BibitemShut {NoStop}%
\bibitem [{\citenamefont {Bastarrachea-Magnani}\ \emph
  {et~al.}(2016)\citenamefont {Bastarrachea-Magnani}, \citenamefont
  {L\'opez-del{-}Carpio}, \citenamefont {Ch\'avez-Carlos}, \citenamefont
  {Lerma-Hern\'andez},\ and\ \citenamefont {Hirsch}}]{Bastarrachea2016PRE}%
  \BibitemOpen
  \bibfield  {author} {\bibinfo {author} {\bibfnamefont {M.~A.}\ \bibnamefont
  {Bastarrachea-Magnani}}, \bibinfo {author} {\bibfnamefont {B.}~\bibnamefont
  {L\'opez-del{-}Carpio}}, \bibinfo {author} {\bibfnamefont {J.}~\bibnamefont
  {Ch\'avez-Carlos}}, \bibinfo {author} {\bibfnamefont {S.}~\bibnamefont
  {Lerma-Hern\'andez}}, \ and\ \bibinfo {author} {\bibfnamefont {J.~G.}\
  \bibnamefont {Hirsch}},\ }\bibfield  {title} {\enquote {\bibinfo {title}
  {Delocalization and quantum chaos in atom-field systems},}\ }\href {\doibase
  10.1103/PhysRevE.93.022215} {\bibfield  {journal} {\bibinfo  {journal} {Phys.
  Rev. E}\ }\textbf {\bibinfo {volume} {93}},\ \bibinfo {pages} {022215}
  (\bibinfo {year} {2016})}\BibitemShut {NoStop}%
\bibitem [{\citenamefont {Hepp}\ and\ \citenamefont
  {Lieb}(1973{\natexlab{b}})}]{Hepp1973b}%
  \BibitemOpen
  \bibfield  {author} {\bibinfo {author} {\bibfnamefont {Klaus}\ \bibnamefont
  {Hepp}}\ and\ \bibinfo {author} {\bibfnamefont {Elliott~H.}\ \bibnamefont
  {Lieb}},\ }\bibfield  {title} {\enquote {\bibinfo {title} {Equilibrium
  statistical mechanics of matter interacting with the quantized radiation
  field},}\ }\href {\doibase 10.1103/PhysRevA.8.2517} {\bibfield  {journal}
  {\bibinfo  {journal} {Phys. Rev. A}\ }\textbf {\bibinfo {volume} {8}},\
  \bibinfo {pages} {2517--2525} (\bibinfo {year}
  {1973}{\natexlab{b}})}\BibitemShut {NoStop}%
\bibitem [{\citenamefont {Emary}\ and\ \citenamefont
  {Brandes}(2003)}]{Emary2003}%
  \BibitemOpen
  \bibfield  {author} {\bibinfo {author} {\bibfnamefont {Clive}\ \bibnamefont
  {Emary}}\ and\ \bibinfo {author} {\bibfnamefont {Tobias}\ \bibnamefont
  {Brandes}},\ }\bibfield  {title} {\enquote {\bibinfo {title} {Chaos and the
  quantum phase transition in the {D}icke model},}\ }\href {\doibase
  10.1103/PhysRevE.67.066203} {\bibfield  {journal} {\bibinfo  {journal} {Phys.
  Rev. E}\ }\textbf {\bibinfo {volume} {67}},\ \bibinfo {pages} {066203}
  (\bibinfo {year} {2003})}\BibitemShut {NoStop}%
\bibitem [{\citenamefont {Ch\'avez-Carlos}\ \emph {et~al.}(2016)\citenamefont
  {Ch\'avez-Carlos}, \citenamefont {Bastarrachea-Magnani}, \citenamefont
  {Lerma-Hern\'andez},\ and\ \citenamefont {Hirsch}}]{Chavez2016}%
  \BibitemOpen
  \bibfield  {author} {\bibinfo {author} {\bibfnamefont {J.}~\bibnamefont
  {Ch\'avez-Carlos}}, \bibinfo {author} {\bibfnamefont {M.~A.}\ \bibnamefont
  {Bastarrachea-Magnani}}, \bibinfo {author} {\bibfnamefont {S.}~\bibnamefont
  {Lerma-Hern\'andez}}, \ and\ \bibinfo {author} {\bibfnamefont {J.~G.}\
  \bibnamefont {Hirsch}},\ }\bibfield  {title} {\enquote {\bibinfo {title}
  {Classical chaos in atom-field systems},}\ }\href {\doibase
  10.1103/PhysRevE.94.022209} {\bibfield  {journal} {\bibinfo  {journal} {Phys.
  Rev. E}\ }\textbf {\bibinfo {volume} {94}},\ \bibinfo {pages} {022209}
  (\bibinfo {year} {2016})}\BibitemShut {NoStop}%
\bibitem [{\citenamefont {de~Aguiar}\ \emph {et~al.}(1991)\citenamefont
  {de~Aguiar}, \citenamefont {Furuya}, \citenamefont {Lewenkopf},\ and\
  \citenamefont {Nemes}}]{Deaguiar1991}%
  \BibitemOpen
  \bibfield  {author} {\bibinfo {author} {\bibfnamefont {M.~A.~M.}\
  \bibnamefont {de~Aguiar}}, \bibinfo {author} {\bibfnamefont {K.}~\bibnamefont
  {Furuya}}, \bibinfo {author} {\bibfnamefont {C.~H.}\ \bibnamefont
  {Lewenkopf}}, \ and\ \bibinfo {author} {\bibfnamefont {M.~C.}\ \bibnamefont
  {Nemes}},\ }\bibfield  {title} {\enquote {\bibinfo {title} {Particle-spin
  coupling in a chaotic system: Localization-delocalization in the {H}usimi
  distributions},}\ }\href {http://stacks.iop.org/0295-5075/15/i=2/a=003}
  {\bibfield  {journal} {\bibinfo  {journal} {EPL (Europhys. Lett.)}\ }\textbf
  {\bibinfo {volume} {15}},\ \bibinfo {pages} {125} (\bibinfo {year}
  {1991})}\BibitemShut {NoStop}%
\bibitem [{\citenamefont {Bastarrachea-Magnani}\ \emph
  {et~al.}(2014{\natexlab{a}})\citenamefont {Bastarrachea-Magnani},
  \citenamefont {Lerma-Hern\'andez},\ and\ \citenamefont
  {Hirsch}}]{Bastarrachea2014a}%
  \BibitemOpen
  \bibfield  {author} {\bibinfo {author} {\bibfnamefont {M.~A.}\ \bibnamefont
  {Bastarrachea-Magnani}}, \bibinfo {author} {\bibfnamefont {S.}~\bibnamefont
  {Lerma-Hern\'andez}}, \ and\ \bibinfo {author} {\bibfnamefont {J.~G.}\
  \bibnamefont {Hirsch}},\ }\bibfield  {title} {\enquote {\bibinfo {title}
  {Comparative quantum and semiclassical analysis of atom-field systems. {I}.
  {D}ensity of states and excited-state quantum phase transitions},}\ }\href
  {\doibase 10.1103/PhysRevA.89.032101} {\bibfield  {journal} {\bibinfo
  {journal} {Phys. Rev. A}\ }\textbf {\bibinfo {volume} {89}},\ \bibinfo
  {pages} {032101} (\bibinfo {year} {2014}{\natexlab{a}})}\BibitemShut
  {NoStop}%
\bibitem [{\citenamefont {Bastarrachea-Magnani}\ \emph
  {et~al.}(2014{\natexlab{b}})\citenamefont {Bastarrachea-Magnani},
  \citenamefont {Lerma-Hern\'andez},\ and\ \citenamefont
  {Hirsch}}]{Bastarrachea2014b}%
  \BibitemOpen
  \bibfield  {author} {\bibinfo {author} {\bibfnamefont {M.~A.}\ \bibnamefont
  {Bastarrachea-Magnani}}, \bibinfo {author} {\bibfnamefont {S.}~\bibnamefont
  {Lerma-Hern\'andez}}, \ and\ \bibinfo {author} {\bibfnamefont {J.~G.}\
  \bibnamefont {Hirsch}},\ }\bibfield  {title} {\enquote {\bibinfo {title}
  {Comparative quantum and semiclassical analysis of atom-field systems. {II}.
  {C}haos and regularity},}\ }\href {\doibase 10.1103/PhysRevA.89.032102}
  {\bibfield  {journal} {\bibinfo  {journal} {Phys. Rev. A}\ }\textbf {\bibinfo
  {volume} {89}},\ \bibinfo {pages} {032102} (\bibinfo {year}
  {2014}{\natexlab{b}})}\BibitemShut {NoStop}%
\bibitem [{\citenamefont {Bastarrachea-Magnani}\ \emph
  {et~al.}(2015)\citenamefont {Bastarrachea-Magnani}, \citenamefont {del
  Carpio}, \citenamefont {Lerma-Hern\'andez},\ and\ \citenamefont
  {Hirsch}}]{Bastarrachea2015}%
  \BibitemOpen
  \bibfield  {author} {\bibinfo {author} {\bibfnamefont {Miguel~Angel}\
  \bibnamefont {Bastarrachea-Magnani}}, \bibinfo {author} {\bibfnamefont
  {Baldemar~L\'opez}\ \bibnamefont {del Carpio}}, \bibinfo {author}
  {\bibfnamefont {Sergio}\ \bibnamefont {Lerma-Hern\'andez}}, \ and\ \bibinfo
  {author} {\bibfnamefont {Jorge~G}\ \bibnamefont {Hirsch}},\ }\bibfield
  {title} {\enquote {\bibinfo {title} {Chaos in the {D}icke model: quantum and
  semiclassical analysis},}\ }\href
  {http://stacks.iop.org/1402-4896/90/i=6/a=068015} {\bibfield  {journal}
  {\bibinfo  {journal} {Phys. Scripta}\ }\textbf {\bibinfo {volume} {90}},\
  \bibinfo {pages} {068015} (\bibinfo {year} {2015})}\BibitemShut {NoStop}%
\bibitem [{\citenamefont {Ribeiro}\ \emph {et~al.}(2006)\citenamefont
  {Ribeiro}, \citenamefont {de~Aguiar},\ and\ \citenamefont
  {de~Toledo~Piza}}]{Ribeiro2006}%
  \BibitemOpen
  \bibfield  {author} {\bibinfo {author} {\bibfnamefont {A.~D.}\ \bibnamefont
  {Ribeiro}}, \bibinfo {author} {\bibfnamefont {M.~A.~M.}\ \bibnamefont
  {de~Aguiar}}, \ and\ \bibinfo {author} {\bibfnamefont {A.~F.~R.}\
  \bibnamefont {de~Toledo~Piza}},\ }\bibfield  {title} {\enquote {\bibinfo
  {title} {The semiclassical coherent state propagator for systems with
  spin},}\ }\href {http://stacks.iop.org/0305-4470/39/i=12/a=016} {\bibfield
  {journal} {\bibinfo  {journal} {J. Phys. A}\ }\textbf {\bibinfo {volume}
  {39}},\ \bibinfo {pages} {3085} (\bibinfo {year} {2006})}\BibitemShut
  {NoStop}%
\bibitem [{\citenamefont {Hillery}\ \emph {et~al.}(1984)\citenamefont
  {Hillery}, \citenamefont {O'Connell}, \citenamefont {Scully},\ and\
  \citenamefont {Wigner}}]{Hillery1984}%
  \BibitemOpen
  \bibfield  {author} {\bibinfo {author} {\bibfnamefont {M.}~\bibnamefont
  {Hillery}}, \bibinfo {author} {\bibfnamefont {R.~F.}\ \bibnamefont
  {O'Connell}}, \bibinfo {author} {\bibfnamefont {M.~O.}\ \bibnamefont
  {Scully}}, \ and\ \bibinfo {author} {\bibfnamefont {E.~P.}\ \bibnamefont
  {Wigner}},\ }\bibfield  {title} {\enquote {\bibinfo {title} {Distribution
  functions in physics: Fundamentals},}\ }\href {\doibase
  10.1016/0370-1573(84)90160-1} {\bibfield  {journal} {\bibinfo  {journal}
  {Phys. Rep.}\ }\textbf {\bibinfo {volume} {106}},\ \bibinfo {pages}
  {121--167} (\bibinfo {year} {1984})}\BibitemShut {NoStop}%
\bibitem [{\citenamefont {Wigner}(1932)}]{Wigner1932}%
  \BibitemOpen
  \bibfield  {author} {\bibinfo {author} {\bibfnamefont {E.}~\bibnamefont
  {Wigner}},\ }\bibfield  {title} {\enquote {\bibinfo {title} {On the quantum
  correction for thermodynamic equilibrium},}\ }\href {\doibase
  10.1103/PhysRev.40.749} {\bibfield  {journal} {\bibinfo  {journal} {Phys.
  Rev.}\ }\textbf {\bibinfo {volume} {40}},\ \bibinfo {pages} {749--759}
  (\bibinfo {year} {1932})}\BibitemShut {NoStop}%
\bibitem [{\citenamefont {Goldberg}\ \emph {et~al.}(2020)\citenamefont
  {Goldberg}, \citenamefont {Klimov}, \citenamefont {Grassl}, \citenamefont
  {Leuchs},\ and\ \citenamefont {S\'anchez-Soto}}]{Goldberg2020}%
  \BibitemOpen
  \bibfield  {author} {\bibinfo {author} {\bibfnamefont {Aaron~Z.}\
  \bibnamefont {Goldberg}}, \bibinfo {author} {\bibfnamefont {Andrei~B.}\
  \bibnamefont {Klimov}}, \bibinfo {author} {\bibfnamefont {Markus}\
  \bibnamefont {Grassl}}, \bibinfo {author} {\bibfnamefont {Gerd}\ \bibnamefont
  {Leuchs}}, \ and\ \bibinfo {author} {\bibfnamefont {Luis~L.}\ \bibnamefont
  {S\'anchez-Soto}},\ }\bibfield  {title} {\enquote {\bibinfo {title} {Extremal
  quantum states},}\ }\href {\doibase 10.1116/5.0025819} {\bibfield  {journal}
  {\bibinfo  {journal} {AVS Quantum Science}\ }\textbf {\bibinfo {volume}
  {2}},\ \bibinfo {pages} {044701} (\bibinfo {year} {2020})}\BibitemShut
  {NoStop}%
\bibitem [{\citenamefont {Gutzwiller}(1971)}]{Gutzwiller1971}%
  \BibitemOpen
  \bibfield  {author} {\bibinfo {author} {\bibfnamefont {Martin~C.}\
  \bibnamefont {Gutzwiller}},\ }\bibfield  {title} {\enquote {\bibinfo {title}
  {Periodic orbits and classical quantization conditions},}\ }\href {\doibase
  10.1063/1.1665596} {\bibfield  {journal} {\bibinfo  {journal} {J. Math.
  Phys.}\ }\textbf {\bibinfo {volume} {12}},\ \bibinfo {pages} {343--358}
  (\bibinfo {year} {1971})}\BibitemShut {NoStop}%
\bibitem [{\citenamefont {Gutzwiller}(1990)}]{Gutzwiller1990book}%
  \BibitemOpen
  \bibfield  {author} {\bibinfo {author} {\bibfnamefont {M.~C.}\ \bibnamefont
  {Gutzwiller}},\ }\href@noop {} {\emph {\bibinfo {title} {Chaos in classical
  and quantum mechanics}}}\ (\bibinfo  {publisher} {Springer-Verlag},\ \bibinfo
  {address} {New York},\ \bibinfo {year} {1990})\BibitemShut {NoStop}%
\bibitem [{\citenamefont {Schreiber}\ and\ \citenamefont
  {Grussbach}(1991)}]{Schreiber1991}%
  \BibitemOpen
  \bibfield  {author} {\bibinfo {author} {\bibfnamefont {Michael}\ \bibnamefont
  {Schreiber}}\ and\ \bibinfo {author} {\bibfnamefont {Heiko}\ \bibnamefont
  {Grussbach}},\ }\bibfield  {title} {\enquote {\bibinfo {title} {Multifractal
  wave functions at the {A}nderson transition},}\ }\href {\doibase
  10.1103/PhysRevLett.67.607} {\bibfield  {journal} {\bibinfo  {journal} {Phys.
  Rev. Lett.}\ }\textbf {\bibinfo {volume} {67}},\ \bibinfo {pages} {607--610}
  (\bibinfo {year} {1991})}\BibitemShut {NoStop}%
\bibitem [{\citenamefont {Mirlin}\ and\ \citenamefont
  {Evers}(2000)}]{Mirlin2000}%
  \BibitemOpen
  \bibfield  {author} {\bibinfo {author} {\bibfnamefont {A.~D.}\ \bibnamefont
  {Mirlin}}\ and\ \bibinfo {author} {\bibfnamefont {F.}~\bibnamefont {Evers}},\
  }\bibfield  {title} {\enquote {\bibinfo {title} {Multifractality and critical
  fluctuations at the {A}nderson transition},}\ }\href {\doibase
  10.1103/PhysRevB.62.7920} {\bibfield  {journal} {\bibinfo  {journal} {Phys.
  Rev. B}\ }\textbf {\bibinfo {volume} {62}},\ \bibinfo {pages} {7920--7933}
  (\bibinfo {year} {2000})}\BibitemShut {NoStop}%
\bibitem [{\citenamefont {Mirlin}(2000)}]{Mirlin2000b}%
  \BibitemOpen
  \bibfield  {author} {\bibinfo {author} {\bibfnamefont {Alexander~D.}\
  \bibnamefont {Mirlin}},\ }\bibfield  {title} {\enquote {\bibinfo {title}
  {Statistics of energy levels and eigenfunctions in disordered systems},}\
  }\href {\doibase 10.1016/S0370-1573(99)00091-5} {\bibfield  {journal}
  {\bibinfo  {journal} {Phys. Rep.}\ }\textbf {\bibinfo {volume} {326}},\
  \bibinfo {pages} {259--382} (\bibinfo {year} {2000})}\BibitemShut {NoStop}%
\bibitem [{\citenamefont {Evers}\ and\ \citenamefont
  {Mirlin}(2008)}]{Evers2008}%
  \BibitemOpen
  \bibfield  {author} {\bibinfo {author} {\bibfnamefont {Ferdinand}\
  \bibnamefont {Evers}}\ and\ \bibinfo {author} {\bibfnamefont {Alexander~D.}\
  \bibnamefont {Mirlin}},\ }\bibfield  {title} {\enquote {\bibinfo {title}
  {Anderson transitions},}\ }\href {\doibase 10.1103/RevModPhys.80.1355}
  {\bibfield  {journal} {\bibinfo  {journal} {Rev. Mod. Phys.}\ }\textbf
  {\bibinfo {volume} {80}},\ \bibinfo {pages} {1355--1417} (\bibinfo {year}
  {2008})}\BibitemShut {NoStop}%
\bibitem [{\citenamefont {Rodriguez}\ \emph {et~al.}(2010)\citenamefont
  {Rodriguez}, \citenamefont {Vasquez}, \citenamefont {Slevin},\ and\
  \citenamefont {R\"omer}}]{Rodriguez2010}%
  \BibitemOpen
  \bibfield  {author} {\bibinfo {author} {\bibfnamefont {Alberto}\ \bibnamefont
  {Rodriguez}}, \bibinfo {author} {\bibfnamefont {Louella~J.}\ \bibnamefont
  {Vasquez}}, \bibinfo {author} {\bibfnamefont {Keith}\ \bibnamefont {Slevin}},
  \ and\ \bibinfo {author} {\bibfnamefont {Rudolf~A.}\ \bibnamefont
  {R\"omer}},\ }\bibfield  {title} {\enquote {\bibinfo {title} {Critical
  parameters from a generalized multifractal analysis at the {A}nderson
  transition},}\ }\href {\doibase 10.1103/PhysRevLett.105.046403} {\bibfield
  {journal} {\bibinfo  {journal} {Phys. Rev. Lett.}\ }\textbf {\bibinfo
  {volume} {105}},\ \bibinfo {pages} {046403} (\bibinfo {year}
  {2010})}\BibitemShut {NoStop}%
\bibitem [{\citenamefont {Martin}\ \emph {et~al.}(2010)\citenamefont {Martin},
  \citenamefont {Garc\'{\i}a-Mata}, \citenamefont {Giraud},\ and\ \citenamefont
  {Georgeot}}]{Martin2010}%
  \BibitemOpen
  \bibfield  {author} {\bibinfo {author} {\bibfnamefont {John}\ \bibnamefont
  {Martin}}, \bibinfo {author} {\bibfnamefont {Ignacio}\ \bibnamefont
  {Garc\'{\i}a-Mata}}, \bibinfo {author} {\bibfnamefont {Olivier}\ \bibnamefont
  {Giraud}}, \ and\ \bibinfo {author} {\bibfnamefont {Bertrand}\ \bibnamefont
  {Georgeot}},\ }\bibfield  {title} {\enquote {\bibinfo {title} {Multifractal
  wave functions of simple quantum maps},}\ }\href {\doibase
  10.1103/PhysRevE.82.046206} {\bibfield  {journal} {\bibinfo  {journal} {Phys.
  Rev. E}\ }\textbf {\bibinfo {volume} {82}},\ \bibinfo {pages} {046206}
  (\bibinfo {year} {2010})}\BibitemShut {NoStop}%
\bibitem [{\citenamefont {Dubertrand}\ \emph {et~al.}(2015)\citenamefont
  {Dubertrand}, \citenamefont {Garc\'{\i}a-Mata}, \citenamefont {Georgeot},
  \citenamefont {Giraud}, \citenamefont {Lemari\'e},\ and\ \citenamefont
  {Martin}}]{Dubertrand2015}%
  \BibitemOpen
  \bibfield  {author} {\bibinfo {author} {\bibfnamefont {R.}~\bibnamefont
  {Dubertrand}}, \bibinfo {author} {\bibfnamefont {I.}~\bibnamefont
  {Garc\'{\i}a-Mata}}, \bibinfo {author} {\bibfnamefont {B.}~\bibnamefont
  {Georgeot}}, \bibinfo {author} {\bibfnamefont {O.}~\bibnamefont {Giraud}},
  \bibinfo {author} {\bibfnamefont {G.}~\bibnamefont {Lemari\'e}}, \ and\
  \bibinfo {author} {\bibfnamefont {J.}~\bibnamefont {Martin}},\ }\bibfield
  {title} {\enquote {\bibinfo {title} {Multifractality of quantum wave
  functions in the presence of perturbations},}\ }\href {\doibase
  10.1103/PhysRevE.92.032914} {\bibfield  {journal} {\bibinfo  {journal} {Phys.
  Rev. E}\ }\textbf {\bibinfo {volume} {92}},\ \bibinfo {pages} {032914}
  (\bibinfo {year} {2015})}\BibitemShut {NoStop}%
\bibitem [{\citenamefont {Nandkishore}\ and\ \citenamefont
  {Huse}(2015)}]{Nandkishore2015}%
  \BibitemOpen
  \bibfield  {author} {\bibinfo {author} {\bibfnamefont {R.}~\bibnamefont
  {Nandkishore}}\ and\ \bibinfo {author} {\bibfnamefont {D.A.}\ \bibnamefont
  {Huse}},\ }\bibfield  {title} {\enquote {\bibinfo {title} {Many-body
  localization and thermalization in quantum statistical mechanics},}\ }\href
  {\doibase 10.1146/annurev-conmatphys-031214-014726} {\bibfield  {journal}
  {\bibinfo  {journal} {Annu. Rev. Condens. Matter Phys.}\ }\textbf {\bibinfo
  {volume} {6}},\ \bibinfo {pages} {15} (\bibinfo {year} {2015})}\BibitemShut
  {NoStop}%
\bibitem [{\citenamefont {Abanin}\ \emph {et~al.}(2019)\citenamefont {Abanin},
  \citenamefont {Altman}, \citenamefont {Bloch},\ and\ \citenamefont
  {Serbyn}}]{Abanin2019}%
  \BibitemOpen
  \bibfield  {author} {\bibinfo {author} {\bibfnamefont {Dmitry~A.}\
  \bibnamefont {Abanin}}, \bibinfo {author} {\bibfnamefont {Ehud}\ \bibnamefont
  {Altman}}, \bibinfo {author} {\bibfnamefont {Immanuel}\ \bibnamefont
  {Bloch}}, \ and\ \bibinfo {author} {\bibfnamefont {Maksym}\ \bibnamefont
  {Serbyn}},\ }\bibfield  {title} {\enquote {\bibinfo {title} {Colloquium:
  Many-body localization, thermalization, and entanglement},}\ }\href {\doibase
  10.1103/RevModPhys.91.021001} {\bibfield  {journal} {\bibinfo  {journal}
  {Rev. Mod. Phys.}\ }\textbf {\bibinfo {volume} {91}},\ \bibinfo {pages}
  {021001} (\bibinfo {year} {2019})}\BibitemShut {NoStop}%
\bibitem [{\citenamefont {Durrett}(2019)}]{Durrett2019}%
  \BibitemOpen
  \bibfield  {author} {\bibinfo {author} {\bibfnamefont {Richard}\ \bibnamefont
  {Durrett}},\ }\href@noop {} {\emph {\bibinfo {title} {Probability: Theory and
  Examples}}},\ \bibinfo {edition} {5th}\ ed.\ (\bibinfo  {publisher}
  {Cambridge University Press},\ \bibinfo {year} {2019})\BibitemShut {NoStop}%
\bibitem [{\citenamefont {Acz\'el}\ and\ \citenamefont
  {Dar\'oczy}(1975)}]{Aczel1975}%
  \BibitemOpen
  \bibfield  {author} {\bibinfo {author} {\bibfnamefont {J.}~\bibnamefont
  {Acz\'el}}\ and\ \bibinfo {author} {\bibfnamefont {Z.}~\bibnamefont
  {Dar\'oczy}},\ }\bibfield  {title} {\enquote {\bibinfo {title} {{R}\'enyi
  entropies},}\ }in\ \href {\doibase 10.1016/S0076-5392(08)62737-X} {\emph
  {\bibinfo {booktitle} {On Measures of Information and their
  Characterizations}}},\ \bibinfo {series} {Mathematics in Science and
  Engineering}, Vol.\ \bibinfo {volume} {115},\ \bibinfo {editor} {edited by\
  \bibinfo {editor} {\bibfnamefont {J.}~\bibnamefont {Acz\'el}}\ and\ \bibinfo
  {editor} {\bibfnamefont {Z.}~\bibnamefont {Dar\'oczy}}}\ (\bibinfo
  {publisher} {Elsevier},\ \bibinfo {year} {1975})\ Chap.~\bibinfo {chapter}
  {5}, pp.\ \bibinfo {pages} {134 -- 172}\BibitemShut {NoStop}%
\bibitem [{\citenamefont {Chen}\ \emph {et~al.}(2008)\citenamefont {Chen},
  \citenamefont {Zhang}, \citenamefont {Liu},\ and\ \citenamefont
  {Wang}}]{Chen2008}%
  \BibitemOpen
  \bibfield  {author} {\bibinfo {author} {\bibfnamefont {Qing-Hu}\ \bibnamefont
  {Chen}}, \bibinfo {author} {\bibfnamefont {Yu-Yu}\ \bibnamefont {Zhang}},
  \bibinfo {author} {\bibfnamefont {Tao}\ \bibnamefont {Liu}}, \ and\ \bibinfo
  {author} {\bibfnamefont {Ke-Lin}\ \bibnamefont {Wang}},\ }\bibfield  {title}
  {\enquote {\bibinfo {title} {Numerically exact solution to the finite-size
  {D}icke model},}\ }\href {\doibase 10.1103/PhysRevA.78.051801} {\bibfield
  {journal} {\bibinfo  {journal} {Phys. Rev. A}\ }\textbf {\bibinfo {volume}
  {78}},\ \bibinfo {pages} {051801} (\bibinfo {year} {2008})}\BibitemShut
  {NoStop}%
\bibitem [{\citenamefont {Lieb}(1978)}]{Lieb1978}%
  \BibitemOpen
  \bibfield  {author} {\bibinfo {author} {\bibfnamefont {Elliott~H.}\
  \bibnamefont {Lieb}},\ }\bibfield  {title} {\enquote {\bibinfo {title} {Proof
  of an entropy conjecture of {W}ehrl},}\ }\href {\doibase 10.1007/bf01940328}
  {\bibfield  {journal} {\bibinfo  {journal} {Comm. in Math. Phys.}\ }\textbf
  {\bibinfo {volume} {62}},\ \bibinfo {pages} {35--41} (\bibinfo {year}
  {1978})}\BibitemShut {NoStop}%
\bibitem [{\citenamefont {Lieb}\ and\ \citenamefont
  {Solovej}(2014)}]{Lieb2014}%
  \BibitemOpen
  \bibfield  {author} {\bibinfo {author} {\bibfnamefont {Elliott~H.}\
  \bibnamefont {Lieb}}\ and\ \bibinfo {author} {\bibfnamefont {Jan~Philip}\
  \bibnamefont {Solovej}},\ }\bibfield  {title} {\enquote {\bibinfo {title}
  {Proof of an entropy conjecture for {B}loch coherent spin states and its
  generalizations},}\ }\href {\doibase 10.1007/s11511-014-0113-6} {\bibfield
  {journal} {\bibinfo  {journal} {Act. Math.}\ }\textbf {\bibinfo {volume}
  {212}},\ \bibinfo {pages} {379--398} (\bibinfo {year} {2014})}\BibitemShut
  {NoStop}%
\bibitem [{\citenamefont {Arecchi}\ \emph {et~al.}(1972)\citenamefont
  {Arecchi}, \citenamefont {Courtens}, \citenamefont {Gilmore},\ and\
  \citenamefont {Thomas}}]{Arecchi1972}%
  \BibitemOpen
  \bibfield  {author} {\bibinfo {author} {\bibfnamefont {F.~T.}\ \bibnamefont
  {Arecchi}}, \bibinfo {author} {\bibfnamefont {Eric}\ \bibnamefont
  {Courtens}}, \bibinfo {author} {\bibfnamefont {Robert}\ \bibnamefont
  {Gilmore}}, \ and\ \bibinfo {author} {\bibfnamefont {Harry}\ \bibnamefont
  {Thomas}},\ }\bibfield  {title} {\enquote {\bibinfo {title} {Atomic coherent
  states in quantum optics},}\ }\href {\doibase 10.1103/PhysRevA.6.2211}
  {\bibfield  {journal} {\bibinfo  {journal} {Phys. Rev. A}\ }\textbf {\bibinfo
  {volume} {6}},\ \bibinfo {pages} {2211--2237} (\bibinfo {year}
  {1972})}\BibitemShut {NoStop}%
\end{thebibliography}%
%%%%%%%%%%%%%%%%%%%%%%%%%%%%%%%%%%%%%%%%%%%%%%%%%%

\end{document}